\documentclass[useAMS,usenatbib]{mn2e}

\usepackage{graphicx}
\usepackage{amsfonts}
\usepackage{bm}

\usepackage{amssymb}
\usepackage{amsmath}
\usepackage[T1]{fontenc}
\usepackage{aecompl} 

\usepackage{arydshln}
\newcommand{\COMPLEX}{\mathbb{C}}
\newcommand{\REAL}{\mathbb{R}}
\newcommand{\II}{\mathbb{I}}

\newcommand{\zz}{\bmath{z}}
\newcommand{\zzc}{\bmath{\bar z}}
\newcommand{\rr}{\bmath{r}}
\newcommand{\rrc}{\bmath{\bar r}}
\newcommand{\vv}{\bmath{v}}

\newcommand{\vecg}{\bmath{g}}
\newcommand{\vecgc}{\bmath{\bar g}}

\newcommand{\mat}[1]{{\bmath{#1}}}

\newcommand{\JJ}{\mat{J}} 
\newcommand{\HH}{\mat{H}} 
\newcommand{\HHa}{\mat{\tilde{H}}} 
\newcommand{\DD}{\mat{D}}
\newcommand{\MM}{\mat{M}}
\newcommand{\RR}{\mat{R}}
\newcommand{\VV}{\mat{V}}

\newcommand{\GG}{\mat{G}}
\newcommand{\JHJ}{\JJ^H\JJ} 

\newcommand{\Matrix}[2]{\left [ \begin{array}{@{}#1@{}}#2\end{array} \right ]}
\newcommand{\Stack}[1]{\begin{array}{@{}c@{}}#1\end{array}}

\newcommand{\AUGx}[1]{\mathbf{\breve{#1}}}
\newcommand{\AUG}[1]{\bmath{\breve{#1}}}

\newcommand{\Zz}{\AUG{z}}
\newcommand{\Gg}{\AUG{g}}
\newcommand{\Rr}{\AUG{r}}
\newcommand{\Dd}{\AUG{d}}
\newcommand{\Vv}{\AUG{v}}
\newcommand{\GGg}{\AUGx{G}}
\newcommand{\RRr}{\AUGx{R}}
\newcommand{\DDd}{\AUGx{D}}
\newcommand{\VVv}{\AUGx{V}}
\newcommand{\TOP}{\mathrm{U}}
\newcommand{\LEFT}{\mathrm{L}}
\newcommand{\RIGHT}{\mathrm{R}}
\newcommand{\UL}{\mathrm{UL}}
\newcommand{\Rop}[1]{\mathcal{R}_{{#1}}}
\newcommand{\Lop}[1]{\mathcal{L}_{{#1}}}

\newcommand{\FigDir}{./}
\newcommand{\COH}{{\sc CohJones}}
\newcommand{\StefCal}{{\sc StefCal}}

\title[Radio interferometric gain calibration as a complex optimization problem]{Radio interferometric gain calibration as a complex optimization problem}

\author[O.M.~Smirnov \& C.~Tasse]{O.M.~Smirnov$^{12}$\thanks{E-mail: o.smirnov@ru.ac.za}, C.~Tasse$^{31}$\\
$^1$Department of Physics and Electronics, Rhodes University, PO Box 94, Grahamstown, 6140 South Africa\\
$^2$SKA South Africa, 3rd Floor, The Park, Park Road, Pinelands, 7405 South Africa\\
$^3$GEPI, Observatoire de Paris, CNRS, Universit\'e Paris Diderot,
5 place Jules Janssen, 92190 Meudon, France}

\numberwithin{equation}{section}

\begin{document}

\date{Accepted 2015 February 24.  Received 2015 February 13; in original form 2014 October 30}

\pagerange{\pageref{firstpage}--\pageref{lastpage}} \pubyear{2014}

\maketitle

\label{firstpage}

\begin{abstract}
Recent developments in optimization theory have extended some traditional algorithms for least-squares optimization of 
real-valued functions (Gauss-Newton, Levenberg-Marquardt, etc.) into the domain of complex functions of a complex 
variable. This employs a formalism called the Wirtinger derivative, and derives a full-complex Jacobian counterpart 
to the conventional real Jacobian. We apply these developments to the problem of radio interferometric gain 
calibration, and show how the general complex Jacobian formalism, when combined with conventional optimization 
approaches, yields a whole new family of calibration algorithms, including those for the polarized and 
direction-dependent gain regime. We further extend the Wirtinger calculus to an operator-based matrix calculus 
for describing the polarized calibration regime. Using approximate matrix 
inversion results in computationally efficient implementations; we show that some recently proposed calibration algorithms 
such as {\sc \StefCal} and peeling can be understood as special cases of this, and place them in the context of the general 
formalism. Finally, we present an implementation and some applied results of {\sc CohJones}, another specialized 
direction-dependent calibration algorithm derived from the formalism.

\end{abstract}

\begin{keywords}
Instrumentation: interferometers, Methods: analytical, Methods: numerical, Techniques: interferometric
\end{keywords}

\section*{Introduction}

In radio interferometry, gain calibration consists of solving for the unknown complex antenna gains,
using a known (prior, or iteratively constructed) model of the sky. Traditional (second generation, or 2GC)
calibration employs an instrumental model with a single direction-independent (DI) gain term (which
can be a scalar complex gain, or $2\times2$ complex-valued {\em Jones matrix}) per antenna, per some 
time/frequency interval. Third-generation (3GC) calibration also addresses direction-dependent (DD) effects, which can be 
represented by independently solvable DD gain terms, or by some parameterized instrumental 
model (e.g. primary beams, pointing offsets, ionospheric screens). Different approaches to 
this have been proposed and implemented, mostly in the framework of the radio interferometry measurement equation
\citep[RIME, see][]{ME1}; \citet{RRIME1,RRIME2,RRIME3} provides a recent overview. In this work we will restrict 
ourselves specifically to calibration of the DI and DD gains terms (the latter in the sense of being solved independently
per direction).

Gain calibration is a non-linear least squares (NLLS) problem, since the noise on observed visibilities is 
almost always Gaussian \citep[though other treatments have been proposed by][]{Kazemi2013a}. Traditional approaches
to NLLS problems involve various gradient-based techniques \citep[for an overview, see][]{Madsen-NLLS}, such as 
Gauss-Newton (GN) and Levenberg-Marquardt (LM). These have been restricted to functions of real variables, 
since complex differentiation can be defined in only a very restricted sense (in particular, $\partial\bar z/\partial z$
does not exist in the usual definition). Gains in radio interferometry are complex variables: the traditional 
way out of this conundrum has been to recast the complex NLLS problem as a real problem by treating the real 
and imaginary parts of the gains as independent real variables.

Recent developments in optimization theory \citep{CR-Calculus,ComplexOpt} have shown that using a formalism 
called the {\em Wirtinger complex derivative} \citep{WirtingerDeriv} allows for a mathematically robust definition of a 
complex gradient operator. This leads to the construction of a {\em complex Jacobian}  $\JJ$, which in turn allows for traditional 
NLLS algorithms to be directly applied to the complex variable case. We summarize these developments and 
introduce basic notation in Sect.~\ref{sec:Wirtinger}. In Sect.~\ref{sec:unpol}, we follow on from \citet{Tasse-cohjones}
to apply this theory to the RIME, and derive complex Jacobians for (unpolarized) DI and DD gain calibration.

In principle, the use of Wirtinger calculus and complex Jacobians ultimately results in the same system of LS 
equations as the real/imaginary approach. It does offer two important advantages: (i) equations with complex variables 
are more compact, and are more natural to derive and analyze than their real/imaginary counterparts, and (ii) the structure of the
complex Jacobian can yield new and valuable insights into the problem. This is graphically illustrated in 
Fig.~\ref{fig:JHJ} (in fact, this figure may be considered the central insight of this paper). Methods such as 
GN and LM hinge around a large matrix -- $\JHJ$ -- with dimensions corresponding to the number of 
free parameters; construction and/or inversion of this matrix is often the dominant algorithmic cost. If $\JHJ$ can
be treated as (perhaps approximately) sparse, these costs can be reduced, often drastically. Figure~\ref{fig:JHJ} 
shows the structure of an example $\JHJ$ matrix for a DD gain calibration problem. The left column row shows versions of
$\JHJ$ constructed via the real/imaginary approach, for four different orderings of the solvable parameters. None of
the orderings yield a matrix that is particularly sparse or easily invertible. The right column shows a 
complex $\JHJ$ for the same orderings. Panel (f) reveals sparsity that is not apparent in the real/imaginary 
approach. This sparsity forms the basis of a new fast DD calibration algorithm discussed later in the paper.

In Sect.~\ref{sec:separability}, we show that different algorithms may be derived by combining 
different sparse approximations to $\JHJ$ with conventional GN and LM methods.  In particular, we show 
that \StefCal, a fast DI calibration algorithm recently proposed by \citet{Stefcal},
can be straightforwardly derived from a diagonal approximation to a complex $\JHJ$. We show that the 
complex Jacobian approach naturally extends to the DD case, and that other sparse approximations yield a whole family
of DD calibration algorithms with different scaling properties. One such algorithm, \COH\ \citep{Tasse-cohjones}, has been implemented 
and successfully applied to simulated LOFAR data: this is discussed in Sect.~\ref{sec:implementations}.

In Sect.~\ref{sec:pol} we extend this approach to the fully polarized case, by developing a Wirtinger-like 
operator calculus in which the polarization problem can be formulated succinctly. This naturally yields 
fully polarized counterparts to the calibration algorithms defined previously. In Sect.~\ref{sec:variations}, 
we discuss other algorithmic variations, and make connections to older DD calibration techniques such as 
peeling \citep{JEN:peeling}. 

While the scope of this work is restricted to LS solutions to the DI and DD gain calibration problem, the potential 
applicability of complex optimization to radio interferometry is perhaps broader. We will return to this in the conclusions.

\begin{table}
\caption{\label{tab:notation}Notation and frequently used symbols}
\begin{tabular}{ll}
\hline
$x$        & scalar value $x$    \\
$\bar{x}$        & complex conjugate    \\
$\bmath{x}$  & vector $\bmath{x}$  \\
$\bmath{X}$  & matrix $\bmath{X}$ \\
$\mathbf{X}$  & vector of $2\times2$ matrices $\mathbf{X}=[\bmath{X}_i]$ (Sect.~\ref{sec:pol})  \\
$\REAL$ & space of real numbers \\
$\COMPLEX$ & space of complex numbers \\
$\II$ & identity matrix \\
$\mathrm{diag}\,\bmath{x}$ & diagonal matrix formed from $\bmath{x}$\\
$||\cdot||_F$ & Frobenius norm \\
$(\cdot)^T$ & transpose \\
$(\cdot)^H$ & Hermitian transpose \\
$\otimes$ & outer product a.k.a. Kronecker product\\
$\bmath{\bar x},\bmath{\bar X}$ & element-by-element complex conjugate of $\bmath{x}$, $\bmath{X}$  \\
$\AUG{x}, \AUGx{X}$ & augmented vectors $\AUG{x} = \Matrix{c}{\bmath{x} \\ \bmath{\bar x}},~~\AUGx{X}=\Matrix{c}{\bmath{X}_i \\ \bmath{X}^H_i}$ \\
$\bmath{X}_\TOP$ & upper half of matrix $\bmath{X}$\\
$\bmath{X}_\LEFT, \bmath{X}_\RIGHT$ & left, right half of matrix $\bmath{X}$ \\
$\bmath{X}_\UL$ & upper left quadrant of matrix $\bmath{X}$\\
& order of operations is $\bmath{X}^Y_\TOP = \bmath{X}^{~~Y}_\TOP = (\bmath{X}_\TOP)^Y$, \\
& or $\bmath{X}^Y_{~\TOP} = (\bmath{X}^Y)_\TOP$ \\
$\bmath{d},\bmath{v},\bmath{r}, \bmath{g}, \bmath{m} $ & data, model, residuals, gains, sky coherency\\
$(\cdot)_k$ & value associated with iteration $k$ \\
$(\cdot)_{p,k}$ & value associated with antenna $p$, iteration $k$ \\
$(\cdot)^{(d)}$ & value associated with direction $d$\\
$\bmath{W}$ & matrix of weights \\
$\JJ_k, \JJ_{k^*}$ & partial, conjugate partial Jacobian at iteration $k$\\
$\JJ$ & full complex Jacobian \\
$\bmath{H}, \bmath{\tilde{H}}$ & $\JHJ$ and its approximation \\
$\mathrm{vec}\,\bmath{X}$ & vectorization operator \\
$\Rop{A}$ & right-multiply by $\bmath{A}$ operator (Sect.~\ref{sec:pol}) \\
$\Lop{A}$ & left-multiply by $\bmath{A}$ operator (Sect.~\ref{sec:pol}) \\
$\delta^i_j$ & Kronecker delta symbol \\
$\Matrix{c@{}c@{}c}{
 A~ & \big |~ & B \\[2pt]
 \hdashline \\[-8pt]
 C~ & \big |~ & D }
$ & matrix blocks \\
$\searrow,\nearrow,\downarrow$ & repeated matrix block \\
\hline

\end{tabular}
\end{table}

\section{Wirtinger calculus \& complex least-squares}
\label{sec:Wirtinger}

The traditional approach to optimizing a function of $n$ complex variables $f(\zz),$ $\zz\in\COMPLEX^n$ is
to treat the real and imaginary parts $\zz=\bmath{x}+i\bmath{y}$ independently, turning $f$ into a function
of $2n$ real variables $f(\bmath{x},\bmath{y})$, and the problem into an optimization over $\REAL^{2n}$.

\citet{CR-Calculus} and \citet{ComplexOpt} propose an alternative approach to the problem based on \citet{WirtingerDeriv} calculus. 
The central idea of Wirtinger calculus is to treat $\zz$ and $\bar{\zz}$ as independent variables, and optimize $f(\zz,\bar{\zz})$
using the {\em Wirtinger derivatives} 
\begin{equation}
\frac{\partial}{\partial z} = \frac{1}{2}\left ( \frac{\partial}{\partial x} - i\frac{\partial}{\partial y} \right),~~
\frac{\partial}{\partial \bar{z}} = \frac{1}{2}\left ( \frac{\partial}{\partial x} + i\frac{\partial}{\partial y} \right),
\end{equation}
where $z=x+iy$. It is easy to see that  
\begin{equation}
\frac{\partial \bar z}{\partial z} = 
\frac{\partial z}{\partial \bar z} = 0,
\end{equation}
i.e. that $\bar z$ ($z$) is treated as constant when taking the derivative with respect to $z$ ($\bar z$). From this 
it is straightforward to define the \emph{complex gradient} operator 
\begin{equation}
\frac{\partial}{\partial^C \zz} = \left [ \frac{\partial}{\partial \zz}, \frac{\partial}{\partial \bar{\zz}} \right ] = \left [ \frac{\partial}{\partial z_1}, \dots, \frac{\partial}{\partial z_n},
\frac{\partial}{\partial \bar{z}_1}, \dots, \frac{\partial}{\partial \bar{z}_n} \right ],
\end{equation}
from which definitions of the complex Jacobian and complex Hessians naturally follow. The authors then show 
that various optimization techniques developed for real functions can be reformulated using 
complex Jacobians and Hessians, and applied to the complex optimization problem. In particular, they generalize 
the Gauss-Newton (GN) and Levenberg-Marquardt (LM) methods for solving the non-linear least squares (NLLS) 
problem\footnote{It should be stressed that Wirtinger calculus can be applied to a broader range of 
optimization problems than just LS.}
\begin{equation}
\label{eq:LSmin}
\min_{\bmath{z}} ||\bmath{r}(\zz,\zzc)||_F,~~~\mathrm{or}~\min_{\bmath{z}} ||\bmath{d}-\bmath{v}(\zz,\zzc)||_F
\end{equation}
where $\bmath{r},\bmath{d},\bmath{v}$ have values in $\COMPLEX^m$, and $||\cdot||_F$ is the Frobenius norm. The latter form 
refers to LS fitting of the parameterized model $\bmath{v}$ to observed data $\bmath{d}$, and is the preferred formulation
in the context of radio interferometry.

Complex NLLS is implemented as follows. Let us formally treat $\zz$ and $\zzc$ as independent variables, 
define an {\em augmented parameter vector} containing both,
\begin{equation}
\Zz = \Matrix{c}{\zz \\ \zzc}
\end{equation}
and designate its value at step $k$ by $\Zz_k$. Then, define
\begin{equation}
\label{eq:Jkrk}
\JJ_k=\frac{\partial \vv}{\partial \zz} (\Zz_k), ~\JJ_{k^*}=\frac{\partial \vv}{\partial \bar{\zz}}(\Zz_k),~
\Rr_k=\Matrix{c}{\rr(\Zz_k)\\\bar{\rr}(\Zz_k)}
\end{equation}
We'll call the $m\times n$ matrices $\JJ_k$ and $\JJ_{k^*}$ the \emph{partial} and \emph{partial conjugate 
Jacobian}\footnote{\citet{ComplexOpt} define the Jacobian via $\partial\rr$ rather than
$\partial\vv$. This yields a Jacobian of the opposite sign, and introduces a minus sign into 
Eqs.~\ref{eq:GN} and \ref{eq:LM}. In this paper we use the $\partial\vv$ convention, as is more common in the
context of radio interferometric calibration.}
respectively, 
and the $2m$-vector $\Rr_k$ the \emph{augmented residual vector}. The \emph{complex Jacobian} 
of the model $\vv$ can then be written (in block matrix form) as
\begin{equation}
\label{eq:JJ}
\JJ = \Matrix{cc}{\JJ_k & \JJ_{k^*} \\ \bar{\JJ}_{k^*} & \bar{\JJ}_{k} },
\end{equation}
with the bottom two blocks being element-by-element conjugated versions of the top two. Note the use of $\bar{\JJ}$ to
indicate element-by-element conjugation -- not to be confused with the Hermitian conjugate which we'll invoke later. 
$\JJ$ is a $2m \times 2n$ matrix. The GN update step is defined as
\begin{equation}
\label{eq:GN}
\delta\Zz = \Matrix{c}{\delta \zz \\ \delta \bar{\zz}} = (\JJ^H \JJ)^{-1}\JJ^H \Rr_k,
\end{equation}

The LM approach is similar, but introduces a {\em damping parameter} $\lambda$:
\begin{equation}
\label{eq:LM}
\delta\Zz = \Matrix{c}{\delta \zz \\ \delta \bar{\zz}} = (\JJ^H \JJ + \lambda\DD)^{-1}\JJ^H \Rr_k,
\end{equation}
where $\DD$ is the diagonalized version of $\JJ^H\JJ$. With $\lambda=0$ this becomes equivalent to GN,
with $\lambda\to\infty$ this corresponds to steepest descent (SD) with ever smaller steps.

Note that while $\delta\zz$ and $\delta \bar{\zz}$ are formally computed independently, the structure of the equations 
is symmetric (since the function being minimized -- the Frobenius norm -- is real and symmetric w.r.t. $\zz$ and $\bar{\zz}$),
which ensures that $\overline{\delta\zz} = \delta \bar{\zz}$. In practice this  redundancy usually means that only half the calculations need to be performed.

\citet{ComplexOpt} show that Eqs.~\ref{eq:GN} and \ref{eq:LM} yield exactly the same system of LS equations as would have 
been produced had we treated $\rr(\zz)$ as a function of real and imaginary parts $\rr(\bmath{x},\bmath{y})$, 
and taken ordinary derivatives in $\REAL^{2n}$. However, the complex Jacobian may be easier and more elegant 
to derive analytically, as we'll see below in the case of radio interferometric calibration.

\section{Scalar (unpolarized) calibration}
\label{sec:unpol}

In this section we will apply the formalism above to the scalar case, i.e. that of unpolarized calibration. 
This will then be extended to the fully polarized case in Sect.~\ref{sec:pol}.

\subsection{Direction-independent calibration}
\label{sec:unpol:DI}

\newcommand{\Na}{N_\mathrm{ant}}
\newcommand{\Nbl}{N_\mathrm{bl}}
\newcommand{\Nd}{N_\mathrm{dir}}

Let us first explore the simplest case of direction-independent (DI) calibration. Consider an interferometer
array of $\Na$ antennas measuring $\Nbl=\Na(\Na-1)/2$ pairwise visibilities. Each antenna pair
$pq$ ($1\leq p<q\leq \Na$) 
measures the visibility\footnote{In principle, the {\em autocorrelation} terms $pp$, corresponding to the
total power in the field, are also measured, and may be incorporated into the equations here. 
It is, however, common practice to omit autocorrelations from the interferometric
calibration problem due to their much higher noise, as well as technical difficulties in modeling the
total intensity contribution. The derivations below are equally valid for $p\leq q$; we use $p<q$ for consistency
with practice.}
\begin{equation}
\label{eq:RIME:unpol}
g_p m_{pq} \bar{g}_q + n_{pq},
\end{equation}
where $m_{pq}$ is the (assumed known) sky coherency, $g_p$ is the (unknown) complex gain parameter 
associated with antenna $p$, and $n_{pq}$ is a complex noise term that is Gaussian with a mean of 0 in the real and 
imaginary parts. The calibration problem then consists of estimating the complex antenna gains $\bmath{g}$ by
minimizing residuals in the LS sense:
\begin{equation}
\label{eq:cal:DI}
\min_{\bmath{g}}\sum_{pq}|r_{pq}|^2,~~~r_{pq}=d_{pq}-g_p m_{pq} \bar{g}_q, 
\end{equation}
where $d_{pq}$ are the observed visibilities. Treating this as a complex optimization problem as per the above, 
let us write out the complex Jacobian. 
With a vector of $\Na$ complex parameters $\bmath{g}$ and $\Nbl$ measurements $d_{pq}$, we'll have a full complex
Jacobian of shape $2\Nbl\times2\Na$. It is conventional
to think of visibilities laid out in a visibility matrix; the normal approach at this stage is to vectorize $d_{pq}$ 
by fixing a numbering convention so as to enumerate all the possible antenna pairs $pq$ ($p<q$) using numbers from 1 to $\Nbl$.
Instead, let us keep using $pq$ as a single ``compound index'', with the implicit understanding that $pq$ in 
subscript corresponds to a single index from 1 to $\Nbl$ using \emph{some} fixed enumeration convention. 
Where necessary, we'll write $pq$ in square brackets (e.g. $a_{[pq],i}$) to emphasize this.

Now consider the corresponding partial Jacobian $\JJ_k$ matrix (Eq.~\ref{eq:Jkrk}). This is of shape $\Nbl\times\Na$. 
Using the Wirtinger derivative, we can write the
partial Jacobian in terms of its value at row $[pq]$ and column $j$ as 
\begin{equation}
[ \JJ_k ]_{[pq],j} = \left \{  
  \begin{array}{ll} 
  m_{pq}\bar{g}_q,& j=p, \\
  0, & \mathrm{otherwise.}
  \end{array}
\right .
\end{equation}

In other words, within each column $j$, $\JJ_k$ is only non-zero at rows corresponding to baselines $[jq]$. We can express 
this more compactly using the Kronecker delta:
\begin{equation}
\label{eq:Jk}
J_k = \overbrace{ \Matrix{c}{ m_{pq}\bar{g}_q\delta^{j}_p } } ^{j=1\dots \Na} \bm{ \}} {\scriptstyle [pq]=1\dots \Nbl~(p<q)}
\end{equation}
Likewise, the conjugate partial Jacobian $J_{k^*}$ may be written as
\begin{equation}
\label{eq:Jbark}
J_{k^*} = \overbrace{ \Matrix{c}{ g_p m_{pq} \delta^{j}_q } } ^{j=1\dots \Na} \bm{ \}} {\scriptstyle [pq]=1\dots \Nbl~(p<q)}
\end{equation}
A specific example is provided in Appendix~\ref{sec:3ant}. The full complex Jacobian (Eq.~\ref{eq:JJ}) then 
becomes, in block matrix notation,
\begin{equation}
\label{eq:JJ:di:basic}
\begin{array}{r@{~}cc@{~}cc}
                & \overbrace{~~~~~~~~}^{j=1\dots \Na} & \overbrace{~~~~~~~~}^{j=1\dots \Na} \\
\JJ = \bigg [ &
  \Stack{ m_{pq}\bar{g}_q\delta^{j}_p \\ \bar{m}_{pq} \bar{g}_p \delta^{j}_q } &
  \Stack{ g_p m_{pq} \delta^{j}_q \\ g_q \bar{m}_{pq} \delta^{j}_p }  
& \bigg ] &
\Stack{ \bm{\}} {\scriptstyle [pq]=1\dots \Nbl}~(p<q) \\ \bm{\}} {\scriptstyle [pq]=1\dots \Nbl}~(p<q) }
\end{array}
\end{equation}
where the $[pq]~(p<q)$ and $j$ subscripts within each block span the full range of $1\dots \Nbl$ and 
$1\dots \Na$. Now, since $d_{pq} = \bar{d}_{qp}$ and $m_{pq} = \bar{m}_{qp}$, we may notice
that the bottom half of the augumented residuals vector $\Rr$ corresponds to the conjugate baselines 
$qp$ ($q>p$):
\begin{equation}
\Rr = \Matrix{c}{ r_{pq} \\ \bar{r}_{pq} } = \Matrix{c}{d_{pq}-g_p m_{pq}\bar{g}_q\\ \bar{d}_{pq}-\bar{g}_p \bar{m}_{pq}g_q} = 
\Matrix{c}{ r_{pq} \\ r_{qp} }~~ 
\Stack{ \bm{\}} \scriptstyle [pq]=1\dots \Nbl~(p<q) \\ \bm{\}} \scriptstyle [pq]=1\dots \Nbl~(p<q) }
\end{equation}
as does the bottom half of $\JJ$ in Eq.~\ref{eq:JJ:di:basic}. Note that we are free to reorder the rows of $\JJ$ and $\Rr$ 
and intermix the normal and conjugate baselines, as this will not affect the LS equations derived at Eq.~\ref{eq:LM}.
This proves most convenient: instead of splitting $\JJ$ and $\Rr$ into 
two vertical blocks with the compound index $[pq]~(p<q)$ running through $\Nbl$ rows within each block, we can treat 
the two blocks as one, with a single compound index $[pq]~(p\ne q)$ running through $2\Nbl$ rows:
\begin{equation}
\label{eq:JJ:di}
\begin{array}{r@{~}cc@{~}cc}
                & \overbrace{~~~~~~~~}^{j=1\dots \Na} & \overbrace{~~~~~~~~}^{j=1\dots \Na} \\
\JJ = \big [ & m_{pq}\bar{g}_q\delta^{j}_p & g_p m_{pq} \delta^{j}_q & \big ],~
\Rr = \big [ r_{pq} \big ] ~~{\bm{\}} \scriptstyle [pq]=1\dots 2\Nbl}
\end{array}
\end{equation}
where for $q>p$, $r_{pq}=\bar{r}_{qp}$ and $m_{pq}=\bar{m}_{qp}$. For clarity, we may adopt the 
following order for enumerating the row index $[pq]$: $12,13,\dots,1n,$ $21,22,\dots,2n,$ $31,32,\dots,3n,$ $\dots,n1,\dots,n\,n-1$. 

Equation~\ref{eq:JJ:3ant} in the Appendix provides an example of $\JJ$ for the 3-antenna case. For
brevity, let us define the shorthand 
\begin{equation}
y_{pq} = m_{pq} \bar{g}_q. 
\end{equation}
We can now write out the 
structure of $\JJ^H\JJ$. This is Hermitian, consisting of four $\Na\times\Na$ blocks:
\begin{equation}
\label{eq:JHJ:DI:ABCD}
\JJ^H\JJ = \Matrix{cc}{\mat{A}&\mat{B}\\\mat{C}&\mat{D}} = \Matrix{cc}{\mat{A}&\mat{B}\\\mat{B}^H&\mat{A}}
\end{equation}
since the value at row $i$, column $j$ of each block is
\begin{eqnarray}
A_{ij} = \sum_{pq} \bar{y}_{pq} y_{pq} \delta^{i}_p \delta^{j}_p &=& 
  \left \{ \begin{array}{cc}
        \sum\limits_{q\ne i} |y_{iq}^2|, & \scriptstyle i=j \\
        0,  & \scriptstyle  i\ne j
  \end{array} \right .\nonumber\\ 
B_{ij} = \sum_{pq} \bar{y}_{pq} \bar{y}_{qp} \delta^{i}_p \delta^{j}_q &=& 
  \left \{ \begin{array}{cc}
      \bar{y}_{ij} \bar{y}_{ji}, & \scriptstyle i\ne j\\
      0, & \scriptstyle i=j
  \end{array} \right .\nonumber\\ 
C_{ij} = \sum_{pq} y_{qp} y_{pq} \delta^{i}_q \delta^{j}_p &=& 
  \bar{B}_{ij} \nonumber\\
D_{ij} = \sum_{pq} y_{pq} \bar{y}_{pq} \delta^{i}_q \delta^{j}_q &=& A_{ij} 
\label{eq:JHJ:DI:ABCD1}
\end{eqnarray}

\newcommand{\JHJblocksFull}[4]{
\Matrix{c@{}c@{}c}{
 #1 & \bigg |~ & #2 \\[10pt]
 \hdashline \\[-8pt]
 #3 & \bigg |~ & #4 }
}

\newcommand{\JHJblocks}[2]{
\Matrix{c@{}c@{}c}{
 #1 & \big |~ & \nearrow^H \\
 \hdashline \\[-8pt]
 #2 & \bigg |~ & \searrow~~ }
}

We then write $\JJ^H\JJ$ in terms of the four $\Na\times\Na$ blocks as:
\begin{equation}
\label{eq:JHJ:DI}
\JJ^H \JJ = 
\JHJblocksFull{
\mathrm{diag} \sum\limits_{q\ne i} |y_{iq}^2| 
}{
  \left \{ 
  \begin{array}{@{}cc@{}}
   \bar{y}_{ij} \bar{y}_{ji}, &{\scriptstyle i\ne j} \\
   0, &{\scriptstyle i=j}
  \end{array} \right .
}{
  \left \{ 
  \begin{array}{@{}cc@{}}
   y_{ij}y_{ji},&{\scriptstyle i\ne j} \\
   0, &{\scriptstyle i=j}
  \end{array} \right . 
}{
  \mathrm{diag} \sum\limits_{q\ne i} |y^2_{iq}| 
}
\end{equation}
Equation~\ref{eq:JHJ:3ant} in the Appendix provides an example of $\JJ^H\JJ$ for the 3-antenna case. 

\newcommand{\yysq}[2]{{y^2_{#1}+y^2_{#2}}}
\newcommand{\bb}[2]{{\bar{y}_{#1#2}\bar{y}_{#2#1}}}
\newcommand{\bbb}[2]{y_{#1#2}y_{#2#1}}

The other component of the LM/GN equations (Eq.~\ref{eq:LM}) is the $\JJ^H\Rr$ term. This will be a column vector of length $2\Na$. We can write this as a stack of two $\Na$-vectors:
\begin{equation}
\label{eq:JHR:DI}
\JJ^H\Rr = \Matrix{c}{ 
\sum\limits_{pq} \bar{y}_{pq} r_{pq} \delta^{i}_p  \\
\sum\limits_{pq} y_{qp} r_{pq} \delta^{i}_q 
} = \Matrix{c}{
\sum\limits_{q\ne i} \bar{y}_{iq} r_{iq}   \\
\sum\limits_{q\ne i} y_{iq} \bar{r}_{iq}  
}
\Stack{
\bm{\big\}} \scriptstyle i=1\dots \Na \\[1ex] 
\bm{\big\}} \scriptstyle i=1\dots \Na
}
\end{equation}
with the second equality established by swapping $p$ and $q$ in the bottom sum, and making use of $r_{pq}=\bar{r}_{qp}$. Clearly, the bottom half of the vector is the conjugate of the top:
\begin{equation}
\label{eq:JHR:DI1}
\JJ^H\Rr = \Matrix{c}{\bmath{c}\\\bar{\bmath{c}}},~~c_i = \sum\limits_{q\ne i} \bar{y}_{iq} r_{iq}.
\end{equation}

\subsection{Computing the parameter update}

Due to the structure of the RIME, we have a particularly elegant way of computing the GN update step.
By analogy with the augmented residuals vector $\Rr$, we can express the data and model visibilities 
as $2\Nbl$-vectors,
using the compound index $[pq]$ $(p\ne q)$:
\begin{equation}
\Dd = [ d_{pq} ],~~\Vv = [ g_p m_{pq} \bar{g}_q ],~~\Rr =\Dd-\Vv
\end{equation}

As noted by \citet{Tasse-cohjones}, we have the wonderful property that
\begin{equation}
\label{eq:v_Jg}
\Vv = \JJ_\LEFT \bmath{g}  = \frac{1}{2}\JJ \Gg,~~~\mathrm{where}~\Gg = \Matrix{c}{\bmath{g} \\ \bmath{\bar g}},
\end{equation}
(where $\bmath{X}_\LEFT$ designates the left half of matrix $\bmath{X}$ -- see Table~\ref{tab:notation} for a summary of notation),
which basically comes about due to the RIME being bilinear with respect to $\bmath{g}$ and $\bmath{\bar g}$. 
Substituting this into the GN update step, and noting that $\bmath{X}(\bmath{Y}_\LEFT) = (\bmath{XY})_\LEFT$, 
we have 
\begin{equation}
\Matrix{c}{\delta\vecg\\\delta\vecgc} = (\JHJ)^{-1} \JJ^H (\Dd - \JJ_\LEFT \vecg) = (\JHJ)^{-1} \JJ^H \Dd - \vecg.
\end{equation}
Consequently, the updated gain values at each iteration can be derived directly from the data, thus obviating the need
for computing residuals. Additionally, since the bottom half of the equations is simply the conjugate of the top, 
we only need to evaluate the top half:
\begin{equation}
\label{eq:ghat}
\vecg_{k+1} = \vecg_k + \delta\vecg = (\JHJ)^{-1}_{~~~\TOP} \JJ^H \Dd,
\end{equation}
where $\bmath{X}_\TOP$ designates the upper half of matrix $\bmath{X}$. 

The derivation above assumes an exact inversion of $\HH=\JHJ$. In practice, this large matrix can be costly to invert, 
so the algorithms below will substitute it with some cheaper-to-invert approximation $\HHa$. Using the approximate
matrix in the GN update equation, we find instead that
\begin{equation}
\label{eq:ghat:approx}
\vecg_{k+1} = \HHa^{-1}_{~~~\TOP} \JJ^H \Dd + (\II - \HHa^{-1}_{~~~\TOP} \HH_\LEFT )\vecg_k,
\end{equation}
which means that when an approximate $\JHJ$ is in use, the shortcut of Eq.~\ref{eq:ghat} only applies when 
\begin{equation}
\label{eq:ghat:precise}
\HHa^{-1}_{~~~\TOP} \HH_\LEFT=\II.
\end{equation}

We will see examples of both conditions below, so it is worth stressing the difference: Eq.~\ref{eq:ghat}
allows us to compute updated solutions directly from the data vector, bypassing the residuals. This is a 
substantial computational shortcut, however, when an approximate inverse for $\HH$ is in use, it 
does not necessarily apply (or at least is not exact). Under the condition of Eq.~\ref{eq:ghat:precise}, however, 
such a shortcut is exact.

\subsection{Time/frequency solution intervals}
\label{sec:unpol:DI:avg}
\label{sec:solution-intervals}

\newcommand{\Ns}{N_s}

A common use case (especially in low-SNR scenarios) is to employ larger solution intervals. 
That is, we measure multiple visibilities per each baseline $pq$, across an interval of timeslots and
frequency channels, then obtain complex gain solutions that are constant across each interval. The 
minimization problem of Eq.~\ref{eq:RIME:unpol} can then be re-written as
\begin{equation}
\label{eq:cal:DI:tf}
\min_{\bmath{g}}\sum_{pqs}|r_{pqs}|^2, 
~~~r_{pqs} = d_{pqs}-g_p m_{pqs} \bar{g}_q, 
\end{equation}
where $s=1,...,N_s$ is a sample index enumerating all the samples within the 
time/frequency solution interval. We can repeat the derivations above using  $[pqs]$ as a
single compound index. Instead of having shape $2\Nbl\times2\Na$, the Jacobian 
will have a shape of $2\Nbl\Ns\times 2\Na$, and the residual vector will have a length of 
$2\Nbl\Ns$. In deriving the $\JHJ$ term, the sums in Eq.~\ref{eq:JHJ:DI:ABCD} must be taken over all $pqs$ 
rather than just $pq$. Defining the usual shorthand of 
$y_{pqs}=m_{pqs}\bar{g}_q$, we then have:
\begin{equation}
\label{eq:JHJ:DI:tf}
\JJ^H \JJ = 
\JHJblocks{
  \mathrm{diag} \sum\limits_{q\ne i,s} |y^2_{iqs}| 
}{
  \left \{ 
  \begin{array}{@{}cc@{}}
   \sum\limits_{s} y_{ijs}y_{jis},&{\scriptstyle i\ne j} \\
   0, &{\scriptstyle i=j}
  \end{array} \right . 
},
\end{equation}
where the symbols $\searrow$ and $\nearrow^H$ represent a copy and a copy-transpose of the appropriate matrix 
block (as per the structure of Eq.~\ref{eq:JHJ:DI:ABCD}). Likewise, the $\JJ^H\Rr$ term can be written as:
\begin{equation}
\label{eq:JHR:DI:tf}
\JJ^H\Rr 
= \Matrix{c}{
\sum\limits_{q\ne i,s} \bar{y}_{iqs} r_{iqs}   \\
 \hdashline \\[-8pt]
\downarrow^H
}.
\end{equation}

\subsection{Weighting}
\label{sec:DI:W}

Although \citet{ComplexOpt} do not mention this explicitly, it is straightforward to incorporate weights into the 
complex LS problem. Equation~\ref{eq:LSmin} is reformulated as
\begin{equation}
\label{eq:LSmin:w}
\min_{\bmath{z}} ||\mat{W} \Rr(\bmath{z},\bmath{\bar z})||_F,
\end{equation}
where $\mat{W}$ is an $M\times M$ weights matrix (usually, the inverse of the data covariance matrix $\mat{C}$). This then propagates into the LM equations as
\begin{equation}
\label{eq:LM:W}
\delta\Zz = (\JJ^H \mat{W} \JJ + \lambda\II)^{-1}\JJ^H \mat{W} \Rr_k.
\end{equation}

Adding weights to Eqs.~\ref{eq:JHJ:DI:tf} and \ref{eq:JHR:DI:tf}, we arrive at the following:
\begin{equation}
\label{eq:JHJ:DI:tfw}
\JJ^H\mat{W}\JJ = 
\JHJblocks{
  \mathrm{diag} \sum\limits_{q\ne i,s} w_{iqs} |y^2_{iqs}|
}{
  \left \{ 
  \begin{array}{@{}c@{,~}c@{}}
   \sum\limits_{s} w_{ijs} y_{ijs}y_{jis}&{\scriptstyle i\ne j} \\
   0 &{\scriptstyle i=j}
  \end{array} \right . 
}
\end{equation}

\begin{equation}
\label{eq:JHR:DI:tfw}
\JJ^H\mat{W}\Rr 
= \Matrix{c}{
\sum\limits_{q\ne i,s} w_{iqs} \bar{y}_{iqs} r_{iqs}   \\
\hdashline\\[-8pt]
\downarrow^H
}.
\end{equation}

\newcommand{\GGd}{\GG^{(d)}}
\newcommand{\GGdH}{\GG^{(d)H}}
\newcommand{\MMd}{\MM^{(d)}}
\newcommand{\YYd}{\YY^{(d)}}
\newcommand{\YYdH}{\YY^{(d)H}}
\newcommand{\YYc}{\YY^{(c)}}
\newcommand{\YYcH}{\YY^{(c)H}}
\newcommand{\ggd}{g^{(d)}}
\newcommand{\ggdH}{\bar{g}^{(d)}}
\newcommand{\ggc}{g^{(c)}}
\newcommand{\ggcH}{\bar{g}^{(c)}}
\newcommand{\mmc}{m^{(c)}}
\newcommand{\mmd}{m^{(d)}}
\newcommand{\mmcH}{\bar{m}^{(c)}}
\newcommand{\mmdH}{\bar{m}^{(d)}}
\newcommand{\yyd}{y^{(d)}}
\newcommand{\yydH}{\bar{y}^{(d)}}
\newcommand{\yyc}{y^{(c)}}
\newcommand{\yycH}{\bar{y}^{(c)}}

\subsection{Direction-dependent calibration}
\label{sec:unpol:DD}

Let us apply the same formalism to the direction-dependent (DD) calibration problem. 
We reformulate the sky model as a sum of $\Nd$ sky components, each with its own DD 
gain. It has been common practice to do DD gain solutions on larger time/frequency intervals than DI 
solutions, both for SNR reasons, and because short intervals lead to under-constrained solutions
and suppression of unmodeled sources. We therefore incorporate solution intervals into the
equations from the beginning. The minimization problem becomes:
\begin{equation}
\label{eq:cal:dd:unpol}
\min_{\bmath{g}}\sum_{pqs}|r_{pqs}|^2, ~~~
r_{pqs} = d_{pqs} - \sum_{d=1}^{\Nd} \ggd_p \mmd_{pqs} \ggdH_q.
\end{equation}
It's obvious that the Jacobian corresponding to this problem is very similar to the one 
in Eq.~\ref{eq:JJ:di}, but instead of having shape $2\Nbl\times2\Na$, this will have a 
shape of $2\Nbl\Ns\times 2\Na\Nd$. We now treat $[pqs]$ and $[jd]$ as compound indices: 
\begin{equation}
\begin{array}{r@{~}cc@{~}cc}
 & \overbrace{~~~~~~~~~~~~}^{\Stack{\scriptstyle j=1\dots \Na\\\scriptstyle d=1\dots \Nd}} & 
   \overbrace{~~~~~~~~~~~~}^{\Stack{\scriptstyle j=1\dots \Na\\\scriptstyle d=1\dots \Nd}} \\
\JJ = \bigg [ &
  \mmd_{pqs}\ggdH_q \delta^j_p & 
  \ggd_p \mmd_{pqs}  \delta^j_q 
\bigg ] &
\bm{\bigg \}}
\begin{array}{l}
\scriptstyle [pq]=1\dots 2\Nbl~(p\ne q)\\ \scriptstyle s=1\dots\Ns
\end{array}
\end{array}
\end{equation}
Every antenna $j$ and direction $d$ will correspond to a column in $\JJ$, but the specific order of the columns 
(corresponding to the order in which we place the $\ggd_p$ elements in the augmented parameter vector $\Gg$)
is completely up to us. 

Consider now the $\JHJ$ product. This will consist of $2\times2$ blocks, each of shape 
$[\Na\Nd]^2$. Let's use $i,c$ to designate the rows within each block, $j,d$ to designate the columns, 
and define $\yyd_{pqs}=\mmd_{pqs}\ggdH_q$. The $\JHJ$ matrix will then have the following block 
structure:
\begin{equation}
\JHJ = \Matrix{cc}{\bmath{A}&\bmath{B}^H\\\bmath{B}&\bmath{A}}
= \JHJblocks{
  \delta^i_j \sum\limits_{q\ne i,s} \yycH_{iqs} \yyd_{iqs} 
}{
  \left \{ 
  \begin{array}{@{}cc@{}}
   \sum\limits_{s} \yyc_{jis} \yyd_{ijs},&{\scriptstyle i\ne j} \\
   0 &{\scriptstyle i=j}
  \end{array} \right . 
},
\label{eq:JHJ:DD:unpol}
\end{equation}
while the $\JJ^H\Rr$ term will be a vector of length $2\Na\Nd$, with the bottom half again being
a conjugate of the top half. Within each half, we can write out the element corresponding to 
antenna $j$, direction $d$: 
\begin{equation}
\label{eq:JHR:DD:unpol}
\JJ^H\Rr = \Matrix{c}{\bmath{c}\\\bar{\bmath{c}}},~~c_{jd} = \sum\limits_{q\ne j,s} 
\yydH_{jqs} r_{jqs}.
\end{equation}

Finally, let us note that the property of Eq.~\ref{eq:v_Jg} also holds for the DD case. It is easy to see that
\begin{equation}
\label{eq:v_Jg:dd}
\Vv = \bigg [ \sum_{d=1}^{\Nd} \ggd_p \mmd_{pqs} \ggdH_q \bigg ] = \JJ_\LEFT \Gg.
\end{equation}
Consequently, the shortcut of Eq.~\ref{eq:ghat} also applies.

\section{Inverting $\JJ^H\JJ$ and separability}
\label{sec:separability}

In principle, implementing one of the flavours of calibration above is ``just'' a matter of  
plugging Eqs.~\ref{eq:JHJ:DI}+\ref{eq:JHR:DI1}, \ref{eq:JHJ:DI:tf}+\ref{eq:JHR:DI:tf},
\ref{eq:JHJ:DI:tfw}+\ref{eq:JHR:DI:tfw} or \ref{eq:JHJ:DD:unpol}+\ref{eq:JHR:DD:unpol} into one the 
algorithms defined in 
Appendix~\ref{sec:algs}. Note, however, that both the GN and LM algorithms hinge around 
inverting a large matrix. This will have a size of $2\Na$ or $2\Na\Nd$ squared, 
for the DI or DD case respectively. With a naive implementation of matrix inversion, 
which scales cubically, algorithmic costs become dominated by the $O(\Na^3)$ or $O(\Na^3\Nd^3)$
cost of inversion.

\begin{figure*}
\begin{center}
\includegraphics[width=.72\textwidth]{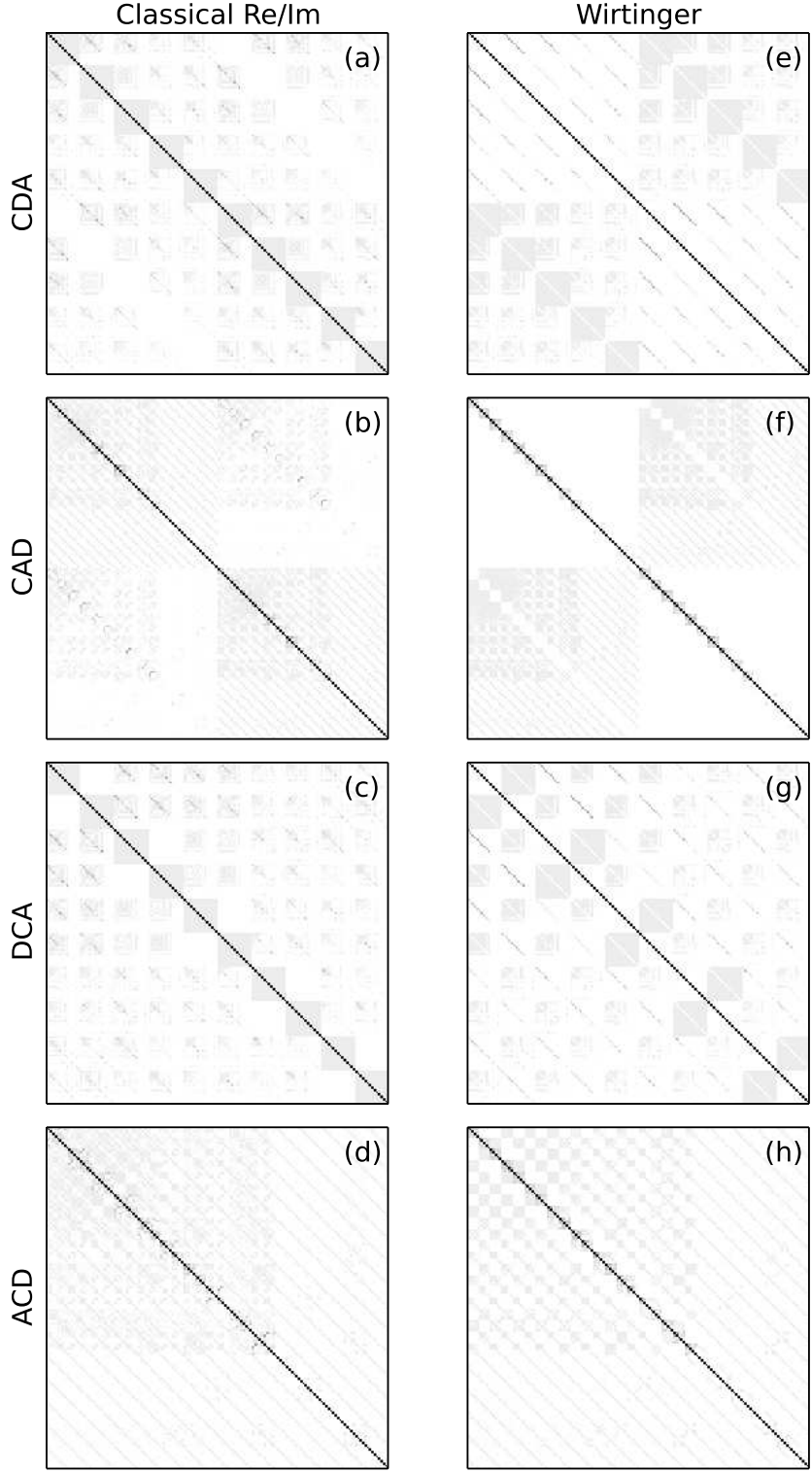}
\caption{\label{fig:JHJ}A graphical representation of $\JHJ$ for a case of 
40 antennas and 5 directions. Each pixel represents the amplitude of a single matrix element.
The left column (a--d) shows conventional real-only Jacobians constructed by taking the partial derivatives w.r.t. 
the real and imaginary parts of the gains. The ordering of the parameters is (a) real/imaginary major, 
direction, antenna minor (i.e. antenna changes fastest); (b) real/imaginary, antenna, direction; (c) direction, 
real/imaginary, antenna; (d) antenna, real/imaginary, direction. The right column (e--h) shows full complex Jacobians with similar parameter ordering (direct/conjugate instead of real/imaginary). Note that panel (f) can also be taken to represent the direction-independent case, if we imagine each $5\times5$ block as one pixel.}
\end{center}
\end{figure*}

In this section we investigate approaches to simplifying the inversion problem by approximating
$\HH=\JJ^H\JJ$ by some form of (block-)diagonal matrix $\HHa$. Such approximation is equivalent to separating
the optimization problem into subsets of parameters that are treated as independent. We will show 
that some of these approximations are similar to or even fully equivalent to previously proposed 
algorithms, while others produce new algorithmic variations.

\subsection{Diagonal approximation and \StefCal}
\label{sec:DI:stefcal}

Let us first consider the DI case. The structure of $\JJ^H\JJ$ in Eq.~\ref{eq:JHJ:DI} suggests that it is diagonally 
dominant (especially for larger $\Na$), as each diagonal element is a coherent sum of $\Na$ amplitude-squared $y$-terms, 
while the off-diagonal elements are either zero or a product of two $y$ terms. This is graphically illustrated in 
Fig.~\ref{fig:JHJ}(f). It is therefore not unreasonable 
to approximate $\JHJ$ with a diagonal matrix for purposes of inversion (or equivalently, making the assumption that 
the problem is separable per antenna):
\begin{equation}
\HHa = \Matrix{cc}{\bmath{A}&0\\0&\bmath{A}}
\end{equation}
This makes the costs of matrix inversion negligible -- $O(\Na)$ operations, as compared to the $O(\Na^2)$ cost 
of computing the diagonal elements of the Jacobian in the first place. The price of using an approximate inverse for 
$\JHJ$ is a less accurate update step, so we can expect to require more iterations before convergence is reached.

Combining this approximation with GN optimization and using Eq.~\ref{eq:ghat:approx}, we find the following expression
for the GN update step:
\begin{equation}
\label{eq:stefcal-matrix}
\vecg_{k+1} = \HHa^{-1}_{\UL} \JJ_\LEFT^{~H} \Dd,
\end{equation}
where $\bmath{X}_\UL$ designates the top left quadrant of matrix $\bmath{X}$. Note that the condition of 
Eq.~\ref{eq:ghat:precise} is met: $\HHa^{-1}_{~~~\TOP} \HH_\LEFT=\bmath{A}^{-1}\bmath{A}=\II$, i.e. the
GN update can be written in terms of $\Dd$. This comes about due to
(i) the off-diagonal blocks of $\HHa$ being null, which masks out the bottom half of $\HH_\LEFT$, and 
(ii) the on-diagonal blocks of $\HHa$ being an exact inverse. In other algorithms suggested below, the second
condition particularly is not the case.

The per-element expression, in the diagonal approximation, is
\begin{equation}
\label{eq:stefcal}
g_{p,k+1} = \big( \sum\limits_{q\ne p} \bar{y}_{pq} y_{pq} \big)^{-1} \sum\limits_{q\ne p} \bar{y}_{pq} d_{pq}.
\end{equation}

Equation~\ref{eq:stefcal} is identical to the update step proposed by \citet{ME4}, and later adopted by 
\citet{Mitchell-RTS} for MWA calibration, and independently derived by \citet{Stefcal} for the \StefCal\ algorithm. 
Note that these authors arrive at the result from a different 
direction, by treating Eq.~\ref{eq:cal:DI} as a function of $\bmath{g}$ only, and completely ignoring 
the conjugate term. 
The resulting complex Jacobian (Eq.~\ref{eq:JJ:di:basic}) then has null off-diagonal blocks, and $\JJ^H\JJ$ becomes 
diagonal.

Interestingly, applying the same idea to LM optimization (Eq.~\ref{eq:LM}), and remembering that $\HHa$ is 
diagonal, we can derive the following update equation instead:
\begin{equation}
\bmath{g}_{k+1} = \frac{\lambda}{1+\lambda}\bmath{g}_k + \frac{1}{1+\lambda} \HHa_{\UL}^{-1} \JJ_\LEFT^{~H} \Dd,
\label{eq:stefcal-lambda-average}
\end{equation}
which for $\lambda=1$ essentially becomes the basic average-update step of \StefCal. We should note that \citet{Stefcal}
empirically find better convergence when Eq.~\ref{eq:stefcal-matrix} is employed for odd $k$, and 
Eq.~\ref{eq:stefcal-lambda-average} for even $k$. In terms of the framework defined here, the basic \StefCal\ algorithm 
can be succinctly described as {\em complex optimization with a 
diagonally-approximated $\JHJ$, using GN for the odd steps, and LM ($\lambda=1$) for the even steps}.

Establishing this equivalence is 
very useful for our purposes, since the convergence properties of \StefCal\ have been thoroughly explored 
by \citet{Stefcal}, and we can therefore hope to apply these lessons here. In particular, these authors have shown 
that a direct application of GN produces very slow (oscillating) convergence, whereas combining GN and LM leads to faster convergence. They also propose a number of variations of the algorithm, all of which are 
directly applicable to the above.

Finally, let us note in passing that the update step of Eq.~\ref{eq:stefcal} is embarrassingly parallel, in the sense 
that the update for each antenna is computed entirely independently.

\subsection{Separability of the direction-dependent case}

Now consider the problem of inverting $\JHJ$ in the DD case. This is a massive matrix, and a brute force 
approach would scale as $O(\Nd^3\Na^3)$. We can, however, adopt a few approximations. Note again that we are 
free to reorder our augmented parameter vector (which contains both $\bmath{g}$ and $\bmath{\bar g}$), 
as long as we reorder the rows and columns of $\JHJ$ accordingly.

Let us consider a number of orderings for $\Gg$:

\begin{itemize}
\item conjugate major, direction, antenna minor (CDA):
\begin{equation}
[g^{(1)}_1 \dots g^{(1)}_{\Na}, g^{(2)}_1 \dots g^{(2)}_{\Na}, g^{(3)}_1 \dots g^{(\Nd)}_{\Na}, 
 \bar{g}^{(1)}_1 \dots \bar{g}^{(1)}_{\Na} \dots ]^T
\end{equation}
\item conjugate, antenna, direction (CAD):
\begin{equation}
[g^{(1)}_1 \dots g^{(\Nd)}_1, g^{(1)}_2 \dots g^{(\Nd)}_2, g^{(1)}_3 \dots g^{(\Nd)}_{\Na}, 
 \bar{g}^{(1)}_1 \dots ]^T
\end{equation}
\item direction, conjugate, antenna (DCA):
\begin{equation}
[g^{(1)}_1 \dots g^{(1)}_{\Na},\bar{g}^{(1)}_1 \dots \bar{g}^{(1)}_{\Na},
g^{(2)}_1 \dots g^{(2)}_{\Na}, \bar{g}^{(2)}_1 \dots ]^T
\end{equation}
\item antenna, conjugate, direction (ACD):
\begin{equation}
[g^{(1)}_1 \dots g^{(\Nd)}_1,\bar{g}^{(1)}_1 \dots \bar{g}^{(\Nd)}_1,
g^{(1)}_2 \dots g^{(\Nd)}_2, \bar{g}^{(1)}_2 \dots ]^T
\end{equation}
\end{itemize}

Figure~\ref{fig:JHJ}(e--h) graphically illustrates the structure of the Jacobian under these orderings.

At this point we may derive a whole family of DD calibration algorithms -- there are many ways to skin a cat. Each 
algorithm is defined by picking an ordering for $\Gg$, then examining the corresponding $\JHJ$ structure 
and specifying an approximate matrix inversion mechanism, then applying GN or LM optimization. Let us now work 
through a couple of examples.

\subsubsection{DCA: separating by direction}
\label{sec:dca}

\newcommand{\JJJ}{\mathcal{J}}

Let us first consider the DCA ordering (Fig.~\ref{fig:JHJ}g). The $\JHJ$ term can be split into $\Nd\times\Nd$ blocks:
\begin{equation}
\label{eq:JHJ:DD:blocked}
\JHJ = \Matrix{ccc}{\JJJ^1_1 & \dots & \JJJ^{\Nd}_1 \\
\vdots & & \vdots \\
\JJJ_{\Nd}^1 & \dots & \JJJ_{\Nd}^{\Nd} },
\end{equation}
where the structure of each $2\Na\times2\Na$ block at row $c$, column $d$, is exactly as 
given by Eq.~\ref{eq:JHJ:DD:unpol} (or by Fig.~\ref{fig:JHJ}f, in miniature). 

The on-diagonal (``same-direction'') blocks $\JJJ^d_d$ will have the same structure as in the DI 
case (Eq.~\ref{eq:JHJ:DI} or Eq.~\ref{eq:JHJ:3ant}). Consider now the off-diagonal (``cross-direction'') 
blocks $\JJJ^d_c$. Their non-zero elements can take one of two forms:
\begin{equation}
  \sum\limits_{q\ne i,s} \yycH_{iqs} \yyd_{iqs} = \sum\limits_{q\ne i} \ggc_q \ggdH_q \sum_s \mmcH_{iqs} \mmd_{iqs}
\end{equation}
or
\begin{equation}
  \sum\limits_{s} \yyc_{jis} \yyd_{ijs} = \ggcH_i \ggdH_j \sum_s \mmcH_{ijs} \mmd_{ijs}
\label{eq:JHJ:DD:crossterms}
\end{equation}
A common element of both is essentially a dot product of sky model components. This is a 
measure of how ``non-orthogonal'' the components are:
\begin{equation}
\label{eq:Xpqcd}
X_{pq}^{(cd)} = \left \langle \bmath{m}_{pq}^\mathrm{(c)},\bmath{m}^\mathrm{(d)}_{pq} \right \rangle = \sum_s \mmc_{pqs} \mmdH_{pqs}.
\end{equation}
We should now note that each model component will typically correspond to a source of limited extent. This can be 
expressed as
\begin{equation}
m_{pqt\nu}^{(d)} = S^{(d)}_{pqt\nu} k^{(d)}_{pqt\nu}, 
\end{equation}
where the term $S$ represents the visibility of that sky model component if placed at phase centre (usually 
only weakly dependent on $t,\nu$ -- in the case of a point source, for example, $S$ is just a constant flux term),
while the term
\begin{equation}
k^{(d)}_{pqt\nu} = e^{-2\pi i (\bmath{u}_{pq}(t)\cdot\bmath{\sigma}_d)\nu/c},~~~\bmath{\sigma}_d = [l_d,m_d,n_d-1]^T,
\end{equation}
represents the phase rotation to direction $\bmath{\sigma}_d$ (where $lmn$ are the corresponding direction cosines), 
given a baseline vector as a function of time $\bmath{u}_{pq}(t)$. We can then approximate the sky model dot product above as
\begin{equation}
X_{pq}^{(cd)} = S^{(c)}_{pq}S^{(d)}_{pq} \sum_s e^{-2\pi i [\bmath{u}_{pq}(t)\cdot(\bmath{\sigma}_c-\bmath{\sigma}_d) ]\nu/c}
\end{equation}

The sum over samples $s$ is essentially just an integral over a complex fringe. We may expect this to be small (i.e. the
sky model components to be more orthogonal) if the directions are well-separated, and also if the sum is taken 
over longer time and frequency intervals. 

If we now assume that the sky model components are orthogonal or near-orthogonal, then 
we may treat the ``cross-direction'' blocks of the $\JHJ$ matrix in Eq.~\ref{eq:JHJ:DD:blocked} as null. The problem is 
then separable by direction, and $\JHJ$ is approximated by a block-diagonal matrix:
\begin{equation}
\label{eq:JHJ:DD:block-diag}
\HHa = \Matrix{ccc}{\JJJ^1_1 &  & 0 \\
& \ddots &  \\
0 & & \JJJ_{\Nd}^{\Nd} },
\end{equation}
The inversion complexity then reduces to $O(\Nd\Na^3)$, which, for large numbers of directions, is a huge improvement on 
$O(\Nd^3\Na^3)$. Either GN and LM optimization may now be applied. 

\newcommand{\JJX}{\bmath{A}}
\newcommand{\JJY}{\bmath{B}}

\subsubsection{COHJONES: separating by antenna}

A complementary approach is to separate the problem by antenna instead. Consider the CAD ordering (Fig.~\ref{fig:JHJ}f). 
The top half of $\JHJ$ then has the following block structure (and the bottom half is its symmetric conjugate):
\begin{equation}
\label{eq:JHJ:DD:blocked-ant}
\HH_\TOP = \Matrix{cccccc}{\JJX^1_1 &  & 0 & \JJY^1_1 & \dots & \JJY^{\Na}_1 \\
 & \ddots &  & \vdots & & \vdots \\
0 &  & \JJX_{\Na}^{\Na} & \JJY_{\Na}^1 & \dots & \JJY_{\Na}^{\Na} },
\end{equation}
that is, its left half is block-diagonal, consisting of $\Nd\times\Nd$ blocks (which follows
from Eq.~\ref{eq:JHJ:DD:unpol}), while its right half consists of elements of the form given by
Eq.~\ref{eq:JHJ:DD:crossterms}. 

By analogy with the \StefCal\ approach, we may assume $\JJY^i_j \approx 0$, i.e. treat the problem as
separable by antenna. The $\HHa$ matrix then becomes block-diagonal, and we only need to compute the 
true matrix inverse of each $\JJX^i_i$. The inversion problem then reduces to $O(\Nd^3\Na)$ in complexity,
and either LM or GN optimization may be applied.

For GN, the update step may be computed in direct analogy to Eq.~\ref{eq:stefcal} (noting that Eq.~\ref{eq:ghat:precise}
holds):
\begin{equation}
\bmath{g}_{k+1} = \HHa_\UL^{-1} \JJ_\LEFT^{~H} \Dd.
\end{equation}
We may note in passing that for LM, the analogy is only approximate:
\begin{equation}
\Gg_{k+1} \approx \frac{\lambda}{1+\lambda}\Gg_k + \frac{1}{1+\lambda} \HHa_\UL^{-1} \JJ_\LEFT^{~H} \Dd,
\end{equation}
since $\HHa$ is only approximately diagonal.

This approach has been implemented as the {\sc CohJones} (complex half-Jacobian optimization for $n$-directional estimation) 
algorithm\footnote{In fact it was the initial development of {\sc CohJones} by \citet{Tasse-cohjones} that directly led to the
present work.}, the results of which applied to simulated data are presented below.


\subsubsection{ALLJONES: Separating all}

Perhaps the biggest simplification available is to start with CDA or CAD ordering, and assume a \StefCal-style diagonal 
approximation for the entirety of $\JHJ$. The matrix then becomes purely diagonal, matrix inversion reduces to $O(\Nd\Na)$ in 
complexity, and algorithmic cost becomes dominated by the $O(\Nd\Na^2)$ process of computing the diagonal elements of 
$\JHJ$. Note that Eq.~\ref{eq:ghat:precise} no longer holds, and the GN update step must be computed via the residuals:
\begin{equation}
\bmath{g}_{k+1} = \bmath{g}_k + \HHa^{-1}_\UL \JJ^{~H}_\LEFT \Rr,
\end{equation}
with the per-element expression being 
\begin{equation}
\label{eq:stefcal:dd:unpol}
g_{p,k+1}^{(d)} = g_{p,k}^{(d)} + 
\bigg( \sum\limits_{q\ne p,s} \bar{y}^{(d)}_{pqs} y^{(d)}_{pqs} \bigg)^{-1}
\sum\limits_{q\ne p,s} \bar{y}^{(d)}_{pqs} r_{pqs}.
\end{equation}

\subsection{Convergence and algorithmic cost}

Evaluating the gain update (e.g. as given by Eq.~\ref{eq:ghat}) involves a number of computational steps:

\begin{enumerate}
\item Computing $\HHa \approx \JHJ$,
\item Inverting $\HHa$,
\item Computing the $\JJ^H\Dd$ vector,
\item Multiplying the result of (ii) and (iii). 
\end{enumerate}

Since each of the algorithms discussed above uses a different sparse approximation for $\JHJ$, each of these 
steps will scale differently (except iii, which is $O(\Na^2\Nd)$ for all algorithms). Table~\ref{tab:costs} 
summarizes the scaling behaviour. 
An additional scaling factor is given by the number of iterations required to converge. This is harder
to quantify. For example, in our experience \citep{OMS-Stefcal}, an ``exact'' (LM) implementation of
DI calibration problem converges in much fewer iterations than \StefCal\ (on the order of a few vs. a few tens), but is much 
slower in terms of ``wall time'' due to the more expensive iterations ($\Na^3$ vs. $\Na$ scaling). This trade-off  
between ``cheap--approximate'' and ``expensive--accurate'' is typical for iterative algorithms. 

\COH\ accounts for interactions between directions, but ignores interactions between antennas. Early experience 
indicates that it converges in a few tens of iterations. {\sc AllJones} uses the most approximative step of all,
ignoring all interactions between parameters. Its convergence behaviour is untested at this time.

It is clear that depending on $\Na$ and $\Nd$, and also on the structure of the problem, there will be 
regimes where one or the other algorithm has a computational advantage. This should be investigated in a future work.

\begin{table}
\caption{\label{tab:costs}The scaling of computational costs for a single iteration of the 
four DD calibration algorithms discussed in the text, broken down by computational step. 
The dominant term(s) in each case are marked by ``$\dagger$''. Not shown is the cost of computing the $\JJ^H\Dd$ 
vector, which is $O(\Na^2\Nd)$ for all algorithms. ``Exact'' refers to a naive implementation of GN or LM with exact 
inversion of the $\JHJ$ term. Scaling laws for DI calibration algorithms may be obtained by assuming $\Nd=1$, in which
case \COH\ or {\sc AllJones} become equivalent to \StefCal. 
}
\begin{tabular}{c|ccc}
\hline
algorithm & $\HHa$ & $\HHa^{-1}$ & multiply \\
\hline
Exact          & $O(\Na^2\Nd^2)$ & $O(\Na^3\Nd^3)^\dagger$ &  $O(\Na^2\Nd^2)$ \\ 
{\sc AllJones} & $O(\Na^2\Nd)^\dagger$   & $O(\Na\Nd)$     &  $O(\Na\Nd)$ \\
\COH           & $O(\Na^2\Nd^2)^\dagger$ & $O(\Na\Nd^3)^\dagger$   &  $O(\Na\Nd^2)$ \\
DCA            & $O(\Na^2\Nd)$   & $O(\Na^3\Nd)^\dagger$   &  $O(\Na^2\Nd)$ \\
\hline
\end{tabular}
\end{table}

\subsection{Smoothing in time and frequency}
\label{sec:DI:smooth}

From physical considerations, we know that gains do not vary arbitrarily in frequency and time. It can
therefore be desirable to impose some sort of smoothness constraint on the solutions, which can improve conditioning, especially
in low-SNR situations. A simple but crude way to do this is use solution intervals (Sect.~\ref{sec:solution-intervals}),
which gives a constant gain solution per interval, but produces non-physical jumps at the edge of each interval.
Other approaches include a posteriori smoothing of solutions done on smaller intervals, as well as various filter-based 
algorithms \citep{tasse-filters}. 

Another way to impose smoothness combines the ideas of solution intervals  (Eq.~\ref{eq:cal:DI:tf}) 
and weighting (Eq.~\ref{eq:LSmin:w}). At every time/frequency sample $t_0,\nu_0$, we can postulate a weighted 
LS problem:
\begin{equation}
\label{eq:cal:DI:smooth}
\min_{\bmath{g}}\sum_{pqt\nu}w(t-t_0,\nu-\nu_0)|r_{pqt\nu}|^2, 
\end{equation}
where $w$ is a smooth weighting kernel that upweighs samples at or near the current sample, and downweighs distant 
samples (e.g., a 2D Gaussian). The solutions for adjacent samples will be very close (since they 
are constrained by practically the same range of data points, with only a smooth change in weights), and the 
degree of smoothness can be controlled by tuning the width of the kernel.

On the face of it this approach is very expensive, since it entails an independent LS solution centred at 
every $t_0,\nu_0$ sample. The diagonal approximation above, however, allows for a particularly elegant and efficient way of 
implementing this in practice. Consider the weighted equations of Eqs.~\ref{eq:JHJ:DI:tfw} and \ref{eq:JHR:DI:tfw}, 
and replace the sample index $s$ by $t,\nu$. Under the diagonal approximation, each parameter update at $t_0,\nu_0$ is 
computed as:
\begin{equation}
\label{eq:stefcal:w}
g_{p,{k+1}}(t_0,\nu_0) = \frac{\sum\limits_{q\ne p,t,\nu} w(t-t_0,\nu-\nu_0) \bar{y}_{pqt\nu} d_{pqt\nu} }
{\sum\limits_{q\ne p,t,\nu} w(t-t_0,\nu-\nu_0) \bar{y}_{pqt\nu} y_{pqt\nu}}.
\end{equation}
Looking at Eq.~\ref{eq:stefcal:w}, it's clear that both sums represent a convolution. If we define two functions of $t,\nu$:
\begin{equation}
\alpha_p(t,\nu) = \sum\limits_{q\ne p} \bar{y}_{pqt\nu} d_{pqt\nu},~~~
\beta_p(t,\nu) = \sum\limits_{q\ne p} \bar{y}_{pqt\nu} y_{pqt\nu},
\end{equation}
then Eq.~\ref{eq:stefcal:w} corresponds to the ratio of two convolutions
\begin{equation}
\label{eq:JHJ:diag:smooth}
g_{p,k+1}(t,\nu) = \frac{w\circ \alpha_p}{w\circ\beta_p},
\end{equation}
sampled over a discrete $t,\nu$ grid. Note that the formulation above also allows for different smoothing kernels per antenna.
Iterating Eq.~\ref{eq:JHJ:diag:smooth} to convergence at every $t,\nu$ slot, we obtain per-antenna arrays of gain solutions 
answering Eq.~\ref{eq:cal:DI:smooth}. These solutions are smooth in frequency and time, with the degree of smoothness 
constrained by the kernel $w$. 

There is a very efficient way of implementing this in practice. Let's assume that $d_{pq}$ and $y_{pq}$ are loaded into memory 
and computed for a large chunk of $t,\nu$ values simultaneously (any practical implementation will probably need to do this anyway, 
if only to take advantage of vectorized math operations on modern CPUs and GPUs). The parameter update step is then also evaluated
for a large chunk of $t,\nu$, as are the $\alpha_p$ and $\beta_p$ terms. We can then take advantage of highly optimized 
implementations of convolution (e.g. via FFTs) that are available on most computing architectures. 

Smoothing may also be trivially incorporated into the {\sc AllJones} algorithm, since its update step (Eq.~\ref{eq:stefcal:dd:unpol}) 
has exactly the same structure. A different smoothing kernel may be employed per direction (for example, directions further from 
the phase centre can employ a narrower kernel). 

Since smoothing involves computing a $\vecg$ value at every $t,\nu$ point, rather than one value per
solution interval, its computational costs are correspondingly higher. To put it another way, using solution intervals of 
size $N_t\times N_\nu$ introduces a savings of $N_t\times N_\nu$ (in terms of the number of invert and multiply steps required, see 
Table~\ref{tab:costs}) over solving the problem at every $t,\nu$ slot; using smoothing foregoes these savings.
In real implementations, this extra cost is mitigated 
by the fact that  the computation given by Eq.~\ref{eq:stefcal:dd:unpol} may be vectorized very efficiently over many $t,\nu$ slots.
However, this vectorization is only straightforward because the matrix inversion in \StefCal\ or {\sc AllJones} reduces to simple 
scalar division.  For \COH\ or DCA this is no longer the case, so while smoothing may be incorporated into these algorithms 
in principle, it is not clear if this can be done efficiently in practice.

\section{The Fully Polarized Case}
\label{sec:pol}

To incorporate polarization, let us start by rewriting the basic RIME of Eq.~\ref{eq:RIME:unpol} using $2\times 2$ matrices 
\citep[a full derivation may be found in][]{RRIME1}:
\begin{equation}
\label{eq:RIME:pol}
\DD_{pq} = \GG_p \MM_{pq} \GG^H_q + \mat{N}_{pq}.
\end{equation}
Here, $\DD_{pq}$ is the \emph{visibility matrix} observed by baseline $pq$, $\MM_{pq}$ is the sky \emph{coherency matrix},
$\GG_p$ is the Jones matrix associated with antenna $p$, and $\mat{N}_{pq}$ is a noise matrix. Quite importantly, 
the visibility and coherency matrices are Hermitian: 
$\DD_{pq}=\DD^H_{qp}$, and $\MM_{pq}=\MM^H_{qp}$. The basic polarization calibration problem can be formulated as
\begin{equation}
\label{eq:cal:DI:pol}
\min_{\{\GG_p\}}\sum_{pq}||\RR_{pq}||_F,~~
\RR_{pq} = \DD_{pq}-\GG_p \MM_{pq} \GG^H_q.
\end{equation}

This is a set of $2\times2$ matrix equations, rather than the vector equations employed in the complex NNLS formalism 
above (Eq.~\ref{eq:LSmin}). In principle, there is a straightforward way of recasting matrix equations into a 
form suitable to  Eq.~\ref{eq:LSmin}: we can vectorize each matrix equation, turning it into an equation on 4-vectors, 
and then derive the complex Jacobian in the usual manner (Eq.~\ref{eq:JJ}). 

In this section we will obtain a more elegant  derivation, by employing an operator calculus where the ``atomic'' 
elements are $2\times2$ matrices rather than scalars. This will allow us to define the Jacobian in a more 
transparent way, as a matrix of linear operators on $2\times2$ matrices. Mathematically, this is completely 
equivalent to vectorizing the problem and applying Wirtinger calculus (each matrix then corresponds to 4 elements 
of the parameter vector, and each operator in the Jacobian becomes a $4\times4$ matrix block). The casual
reader may simply take the postulates of the following section on faith -- in particular, that the operator calculus 
approach is completely equivalent to using $4\times4$ matrices. A rigorous formal footing to this is given 
in Appendix~\ref{sec:opcalculus}.


\newcommand{\VEC}[1]{\mathrm{vec}\,{#1}}
\newcommand{\VECINV}[1]{\mathrm{vec}^{-1}\,{#1}}

\subsection{Matrix operators and derivatives}

By \emph{matrix operator}, we shall refer to any function $\mathcal{F}$ that maps a $2\times2$ complex matrix to 
another such matrix:
\begin{equation}
\mathcal{F}:\COMPLEX^{2\times2} \to \COMPLEX^{2\times2}.
\end{equation}
When the operator $\mathcal{F}$ is applied to matrix $\mat{X}$, we'll write the result 
as $\mat{Y}=\mathcal{F}\mat{X}$, or $\mathcal{F}[\mat{X}]$ if we need to avoid ambiguity.

If we fix a complex matrix $\mat{A}$, then two interesting (and linear) matrix operators are
right-multiply by $\mat{A}$, and left-multiply by $\mat{A}$:
\begin{equation}
\begin{array}{l@{~}l}
\Rop{\mat{A}}\mat{X} &= \mat{XA} \\
\Lop{\mat{A}}\mat{X} &= \mat{AX} \\
\end{array}
\end{equation}
Appendix~\ref{sec:opcalculus} formally shows that all linear matrix operators, including $\Rop{\mat{A}}$ and $\Lop{\mat{A}}$,
can be represented as multiplication of 4-vectors by $4\times4$ matrices. 

Just from the operator definitions, it is trivial to see that
\begin{equation}
\label{eq:RAB:LAB}
\Lop{\mat{A}}\Lop{\mat{B}} = \Lop{\mat{AB}},~~~\Rop{\mat{A}}\Rop{\mat{B}} = \Rop{\mat{BA}},~~~[\Rop{\mat{A}}]^{-1} = \Rop{\mat{A^{-1}}}
\end{equation}



Consider now a matrix-valued function of $n$ matrices and their Hermitian transposes
\begin{equation}
\mat{F}(\mat{G}_1\dots\mat{G}_n,\mat{G}^H_1\dots\mat{G}^H_n),
\end{equation}
and think what a consistent definition for the partial matrix derivative $\partial\mat{F}/\partial\mat{G}_i$ 
would need be. A partial derivative at some fixed point $\GGg_0 = (\mat{G}_1\dots\mat{G}_n,\mat{G}^H_1\dots\mat{G}^H_n)$
is a local linear approximation to $\mat{F}$, i.e. a linear function mapping an increment 
in an argument $\Delta\mat{G}_i$ to an increment in the function value $\Delta\mat{F}$. In other words, the partial 
derivative is a linear matrix operator. Designating this operator as $\mathcal{D}=\partial\mat{F}/\partial\mat{G}_i$, we 
can write the approximation as:
\begin{equation}
\mat{F}(...,\mat{G}_i+\Delta\mat{G}_i,...) - \mat{F}(...,\mat{G}_i,...) \approx \mathcal{D} \Delta\mat{G}_i.
\end{equation}
Obviously, the linear operator that best approximates a given linear operator is the operator itself, so we necessarily have
\begin{equation}
\frac{\partial(\mat{GA})}{\partial{\mat{G}}} = \Rop{A},~~~~\frac{\partial(\mat{AG^H})}{\partial{\mat{G^H}}} = \Lop{A}.
\end{equation}

Appendix~\ref{sec:opcalculus} puts this on a formal footing, by providing formal definitions of \emph{Wirtinger matrix derivatives}
\begin{equation}
\frac{\partial\mat{F}}{\partial{\mat{G}}}, ~~
\frac{\partial\mat{F}}{\partial{\mat{G}^H}} 
\end{equation}
that are completely equivalent to the partial complex Jacobians defined earlier. 

Note that this calculus also offers a natural way of taking more
complicated matrix derivatives (that is, for more elaborate versions of the RIME). For example,
\begin{equation}
\frac{\partial(\mat{AGB})}{\partial{\mat{G}}} = \Lop{A}\Rop{B},
\end{equation}
which is a straightforward manifestation of the chain rule: $\mat{AGB}=\mat{A}(\mat{GB})$.

\subsection{Complex Jacobians for the polarized RIME}

Let us now apply this operator calculus to Eq.~\ref{eq:cal:DI:pol}. Taking the derivatives, we have:
\begin{equation}
\label{eq:dR:dG}
\frac{\partial\VV_{pq}}{\partial\GG_p} = \Rop{\MM_{pq}\GG^H_q},\mathrm{~~and~~}
\frac{\partial\VV_{pq}}{\partial\GG^H_q} = \Lop{\GG_p \MM_{pq}}.
\end{equation}
If we now stack all the gain matrices into one augmented ``vector of matrices'':
\begin{equation}
\label{eq:GGvec}
\GGg = [ \GG_1,\dots,\GG_{\scriptstyle \Na},\GG^H_1,\dots,\GG^H_{\scriptstyle \Na} ]^T,
\end{equation}
then we may construct the top half of the full complex Jacobian operator in full analogy with the 
derivation of Eq.~\ref{eq:JJ:di:basic}. We'll use the same ``compound index'' convention for $pq$. That is, 
$[pq]$ will represent a single index running through $M$ values (i.e. enumerating all combinations of $p<q$).
\begin{equation}
\label{eq:JJtop}
\begin{array}{r@{~}cc@{~}cc}
  & \overbrace{~~~~~~~~}^{j=1\dots \Na} & \overbrace{~~~~~~~~}^{j=1\dots \Na} \\

\JJ_\TOP = \big [ & 
\Rop{ \MM_{pq}\GG^H_q }\delta^j_p & 
\Lop{ \GG_p \MM_{pq}  }\delta^j_q 
& \big ] ~~ \bm{\}} \scriptstyle [pq]=1\dots \Nbl~(p<q)
%
\end{array}
\end{equation}

Note that there are two fully equivalent ways to read the above equation. In operator notation, it specifies a linear 
operator $\JJ_\TOP$ that maps a $2\Na$-vector of $2\times2$ matrices to an $\Nbl$-vector of $2\times2$ matrices. In
conventional matrix notation (Appendix~\ref{sec:opcalculus}), $\JJ_\TOP$ is just a $4\Nbl\times8\Na$ matrix; 
the above equation then specifies the structure of this matrix in terms of $4\times4$ blocks, where each block 
is the matrix equivalent of the appropriate $\mathcal{R}$ or $\mathcal{L}$ operator.

Consider now the bottom half of the Jacobian. In Eq.~\ref{eq:JJ}, this corresponds to the derivatives of the conjugate
residual vector $\rrc_k$, and can be constructed by conjugating and mirroring $\JJ_\TOP$. Let us modify this construction 
by taking the derivative of the Hermitian transpose of the residuals instead. Note that substituting the Hermitian transpose 
for element-by-element conjugation corresponds to a simple reordering of some rows in the conjugate residual vector (i.e.
reordering of the LS equations), which we are always free to do. Let us then construct the augmented residual vector of matrices as:
\begin{equation}
\RRr = 
\Matrix{c}{
  \RR_{pq} \\[1ex] 
  \RR^H_{pq} 
} 
~~ 
\Stack{ 
\bm{\}} \scriptstyle [pq]=1\dots \Nbl~(p<q) \\[1ex] 
\bm{\}} \scriptstyle [pq]=1\dots \Nbl~(p<q) 
}
\label{eq:RRtranspose}
\end{equation}
Now, since $\VV^H_{pq}=\GG_q \MM^H_{pq} \GG^H_p,$ we have
\begin{equation}
\label{eq:dRH:dG}
\frac{\partial\VV^H_{pq}}{\partial\GG_q} = \Rop{\MM^H_{pq}\GG^H_p},\mathrm{~~and~~}
\frac{\partial\VV^H_{pq}}{\partial\GG^H_p} = \Lop{\GG_q \MM^H_{pq}},
\end{equation}
and we may write out the full complex Jacobian as
\begin{equation}
\JJ = - \Matrix{cc}{ 
\Rop{ \MM_{pq}\GG^H_q }\delta^j_p & 
\Lop{ \GG_p \MM_{pq}  }\delta^j_q \\
\Rop{ \MM^H_{pq} \GG^H_p } \delta^j_q & 
\Lop{ \GG_q \MM^H_{pq}  } \delta^j_p  
}~~ 
\Stack{ 
\bm{\}} \scriptstyle [pq]=1\dots \Nbl~(p<q) \\[1ex] 
\bm{\}} \scriptstyle [pq]=1\dots \Nbl~(p<q) 
}
\end{equation}

We may now make exactly the same observation as we did to derive Eq.~\ref{eq:JJ:di}, and rewrite both $\JJ$ and $\RRr$ in terms of 
a single row block. The $pq$ index will now run through $2\Nbl$ values (i.e. enumerating all combinations of $p\ne q$):
\begin{equation}
\label{eq:JJ:pol}
\JJ = \Matrix{cc}{ 
\Rop{ \MM_{pq}\GG^H_q }\delta^j_p & 
\Lop{ \GG_p \MM_{pq}  }\delta^j_q \\
}~~ 
\bm{\big\}} \scriptstyle [pq]=1\dots 2\Nbl~(p\ne q)\\ 
\end{equation}
and 
\begin{equation}
\RRr = \Matrix{c}{ \mat{R}_{pq}}~~ 
\bm{\}} \scriptstyle [pq]=1\dots 2\Nbl~(p\ne q)\\ 
\end{equation}
This is in complete analogy to the derivations of the unpolarized case. For compactness, let us now define
\newcommand{\YY}{\mat{Y}}
\newcommand{\ZZ}{\mat{Z}}
\begin{equation}
\YY_{pq} = \MM_{pq} \GG^H_q,~~\YY_{qp} = \MM^H_{pq} \GG^H_p
\end{equation}
noting that 
\begin{equation}
\YY^H_{pq} = \GG_q \MM^H_{pq},~~\YY_{qp}^H = \GG_p \MM_{pq}.
\end{equation}
Employing Eq.~\ref{eq:RH:LH}, the $\JJ$ and $\JJ^H$ terms can be written as
\begin{equation}
\JJ^H = \Matrix{c}{
\Rop{ Y^H_{pq} }\delta^i_p \\[4pt]
\hdashline\\[-10pt]
\Lop{ Y_{qp}  }\delta^i_q \\
},~~~ 
\JJ = \Matrix{cc}{ 
\Rop{ Y_{pq} }\delta^j_p & 
\Lop{ Y^H_{qp}  }\delta^j_q \\
}~~ 
\end{equation}

We can now write out the $\JHJ$ term, still expressed in terms of operators, as:
\begin{equation}
\label{eq:JHJ:pol}
\JHJ = \JHJblocksFull{
  \mathrm{diag}\sum\limits_{q\ne i} \Rop{Y_{iq} Y^H_{iq}} 
}{
  \left \{ 
  \begin{array}{@{}cc@{}}
   \Rop{ Y^H_{ij}  } \Lop{Y^H_{ji}},&{\scriptstyle i\ne j} \\
   0, &{\scriptstyle i=j}
  \end{array} \right . 
}{
  \left \{ 
  \begin{array}{@{}cc@{}}
   \Lop{ Y_{ij}  } \Rop{ Y_{ji} },&{\scriptstyle i\ne j} \\
   0, &{\scriptstyle i=j}
  \end{array} \right . 
}{
  \mathrm{diag}\sum\limits_{q\ne i} \Lop{Y_{iq} Y^H_{iq}} 
}
\end{equation}
(Note that this makes use of the property $\Rop{\mat{A}}\Rop{\mat{B}}=\Rop{\mat{BA}}$ and 
$\Lop{\mat{A}}\Lop{\mat{B}}=\Lop{\mat{AB}}$.) Compare this result to Eq.~\ref{eq:JHJ:DI}.

As for the $\JJ^H\RRr$ term, we can directly apply the linear operators appearing in $\JJ^H$ 
to the matrices in $\RRr$. This results in the following vector of $2\times2$ matrices:
\begin{equation}
\JJ^H\RRr = \Matrix{c}{ 
\sum\limits_{pq} \YY^H_{pq} \RR_{pq} \delta^{i}_p  \\
\hdashline\\[-8pt]
\sum\limits_{pq} \RR_{pq} \YY_{qp} \delta^{i}_q 
} = \Matrix{c}{
\sum\limits_{q\ne i} \YY^H_{iq} \RR_{iq} \\
\hdashline\\[-8pt]
\downarrow^H
},
\end{equation}
where the second equality is established by swapping the $p$ and $q$ indices. Unsurprisingly, and by analogy with 
Eq.~\ref{eq:JHR:DI1}, the bottom half of the vector is Hermitian with respect to the top.

\subsection{Parameter updates and the diagonal approximation}

The relation of Eq.~\ref{eq:v_Jg} also apply in the fully-polarized case. It is easy to see that if we define the augmented 
data and model vectors of matrices as
\begin{equation}
\DDd = [ \DD_{pq} ],~~\VVv = [ \GG_p \MM_{pq} \GG_q^H ],
\end{equation}
then $\VVv = \JJ_\LEFT \GGg$ holds, and the GN update step can be written as
\begin{equation}
\label{eq:ghat:matrix}
\GGg_{k+1} = (\JHJ)^{-1} \JJ^H \DDd.
\end{equation}
By analogy with Eq.~\ref{eq:ghat}, this equation also holds when $\JHJ$ is approximated, but only if the condition of 
Eq.~\ref{eq:ghat:precise} is met.

To actually implement GN or LM optimization, we still need to invert the operator represented by the $\JHJ$ matrix in 
Eq.~\ref{eq:JHJ:pol}. We have two options here. 

The brute-force numerical approach is to substitute the 
$4\times4$ matrix forms of the $\mathcal{R}$ and $\mathcal{L}$ operators (Eqs.~\ref{eq:RA4x4} and \ref{eq:LA4x4}) into the 
equation, thus resulting in conventional matrix,
and then do a straightforward matrix inversion. 

The second option is to use an analogue of the diagonal approximation described in Sect.~\ref{sec:DI:stefcal}. If we 
neglect the off-diagonal operators of Eq.~\ref{eq:JHJ:pol}, the operator form of the matrix is diagonal, i.e. the problem 
is again treated as being separable per antenna. As for the operators on the diagonal, they are trivially invertible as per 
Eq.~\ref{eq:RAB:LAB}. We can therefore directly invert the operator form
of $\JHJ$ and apply it to $\JJ^H\RRr$, thus arriving at a simple per-antenna equation  for the GN update step:
\begin{equation}
\label{eq:update:DI:diag:pol}
\GG_{p,k+1} =  
\left [ \sum\limits_{q\ne p} \YY^H_{pq} \DD_{pq} \right ] 
\left [ \sum\limits_{q\ne p} \YY_{pq} \YY^H_{pq}  \right ]^{-1}
\end{equation}
This is, once again, equivalent to the polarized {\sc \StefCal} update step proposed by \citet{OMS-Stefcal} and \citet{Stefcal}.


\subsection{Polarized direction-dependent calibration}

Let us now briefly address the fully-polarized DD case. This can be done by direct analogy with
Sect.~\ref{sec:unpol:DD}, using the operator calculus developed above. As a result, we arrive at the
following expression for $\JHJ$:
\begin{equation}
\label{eq:JHJ:DD:pol}
  \JHJblocksFull{
  \delta^i_j \sum\limits_{q\ne i,s} \Rop{\YYd_{iqs} \YYcH_{iqs}} 
  }{
  \left \{ 
  \begin{array}{@{}c}
   \sum\limits_{s} \Rop{ \YYcH_{ijs}  } \Lop{\YYdH_{jis}} \\
   0
  \end{array} \right . 
  }{
  \left \{ 
  \begin{array}{@{}cc@{}}
   \sum\limits_{s} \Lop{ \YYc_{ijs}  } \Rop{\YYd_{jis}},&{\scriptstyle i\ne j} \\
   0 &{\scriptstyle i=j}
  \end{array} \right . 
  }{ 
  \delta^i_j \sum\limits_{q\ne i,s} \Lop{\YYc_{iqs} \YYdH_{iqs}} 
  }
\end{equation}
using the normal shorthand of $\YYd_{pqs} = \MMd_{pqs} \GGdH_q$. The $\JJ^H\RRr$ term is then
\newcommand{\CCC}{\mathcal{C}}
\begin{equation}
\label{eq:JHR:DD:pol}
\JJ^H\RRr = \Matrix{c}{
\sum\limits_{q\ne i,s} \YYdH_{iqs} \RR_{iqs} \\
\hdashline\\[-8pt]
\downarrow^H}.
\end{equation}
All the separability considerations of Sect.~\ref{sec:separability} now apply, and polarized versions of the 
algorithms referenced therein may be reformulated for the fully polarized case. For example:

\begin{itemize} 
\item If we assume separability by both direction and antenna, as in the {\sc AllJones} algorithm, then
the $\HHa$ matrix is fully diagonal in operator form, and the GN update step can be computed as
\begin{equation}
\label{eq:update:DD:diag:pol}
\delta \GGd_{p,k+1} = 
\left [ \sum\limits_{q\ne p,s} \YYdH_{pqs} \RR_{pqs} \right ]
\left [ \sum\limits_{q\ne p,s} \YYd_{pqs} \YYdH_{pqs}  \right ]^{-1}.
\end{equation}
Note that in this case (as in the unpolarized {\sc AllJones} version) the condition of Eq.~\ref{eq:ghat:precise}
is not met, so we must use the residuals and compute $\delta\GG$.

\item If we only assume separability by antenna, as in the {\sc CohJones} algorithm, then the $\HHa$ matrix 
becomes $4\Nd\times4\Nd$-block-diagonal, and may be inverted exactly at a cost of $O(\Nd^3\Na)$. The condition 
of Eq.~\ref{eq:ghat:precise} is met.
\end{itemize}

It is also straightforward to add weights and/or sliding window averaging to this formulation, as per 
Sect.~\ref{sec:DI:W} and \ref{sec:DI:smooth}.

Equations~\ref{eq:JHJ:DD:pol}--\ref{eq:JHR:DD:pol} can be considered the principal result of this work.
They provide the necessary ingredients for implementing GN or LM methods for DD calibration, treating it as a 
fully complex optimization problem. The equations may be combined and approximated in different 
ways to produce different types of calibration algorithms. 

Another interesting note is that Eq.~\ref{eq:update:DD:diag:pol} and its ilk are embarrassingly parallel, since the update step is 
completely separated by direction and antenna. This makes it particularly well-suited to implementation on massively 
parallel architectures such as GPUs.

\section{Other DD algorithmic variations}
\label{sec:variations}

The mathematical framework developed above (in particular, Eqs.~\ref{eq:JHJ:DD:pol}--\ref{eq:JHR:DD:pol}) provides
a general description of the polarized DD calibration problem. Practical implementations of this hinge around inversion of
a very large $\JHJ$ matrix. The family of algorithms proposed in Sect.~\ref{sec:separability} takes different approaches
to approximating this inversion. Their convergence properties are not yet well-understood; however we may note that 
the {\sc \StefCal} algorithm naturally emerges from this formulation as a specific case, and its convergence has been 
established by \citet{Stefcal}. This is encouraging, but ought not be treated as anything more than a strong pointer for 
the DD case. It is therefore well worth exploring other approximations to the problem. In this section we map out a few 
such options.

\subsection{Feed forward}
\label{sec:feed-forward}

\citet{Stefcal-URSI} propose variants of the \StefCal\ algorithm (``2-basic'' and ``2-relax'') where the results of the 
update step (Eq.~\ref{eq:update:DI:diag:pol}, in essence) are computed sequentially per antenna, with updated
values for $\GG_1\dots\GG_{k-1}$ fed forward into the equations for $\GG_k$ (via the appropriate $\YY$ terms). This is shown to 
substantially improve convergence, at the cost of sacrificing the embarrassing parallelism by antenna. This technique 
is directly applicable to both the {\sc AllJones} and {\sc CohJones} algorithms. 

The {\sc CohJones} algorithm considers all directions simultaneously, but could still implement feed-forward by antenna.
The {\sc AllJones} algorithm (Eq.~\ref{eq:update:DD:diag:pol}) could implement feed-forward by both antenna (via $\YY$) and 
by direction -- by recomputing the residuals $\RR$ to take into account the updated solutions for $\GG^{(1)}\dots\GG^{(d-1)}$ before
evaluating the solution for $\GG^{(d)}$.  The optimal order for this, as well as whether in practice this actually 
improves convergence to justify the extra complexity, is an open issue that remains to be investigated.

\subsection{Triangular approximation}

The main idea of feed-forward is to take into account solutions for antennas (and/or directions) $1,...,k-1$ when computing the
solution for $k$. A related approach is to approximate the $\JHJ$ matrix as block-triangular:
\begin{equation}
\label{eq:JHJ:DD:block-triag}
\HHa = \Matrix{cccc}{
\JJJ^1_1 & 0 & \cdots & 0 \\
\JJJ^1_2 & \JJJ^2_2 & \cdots & 0 \\
\vdots & & \ddots &  \vdots \\
\JJJ_{N}^1 & \JJJ_{N}^2 & \cdots & \JJJ_{N}^{N} },
\end{equation}
\newcommand{\KKK}{\mathcal{K}}
The inverse of this is also block triangular:
\begin{equation}
\label{eq:JHJ:DD:block-triag-inv}
\HHa^{-1} = \Matrix{cccc}{
\KKK^1_1 & 0 & \cdots & 0 \\
\KKK^1_2 & \KKK^2_2 & \cdots & 0 \\
\vdots & & \ddots &  \vdots \\
\KKK_{N}^1 & \KKK_{N}^2 & \cdots & \KKK_{N}^{N} },
\end{equation}
which can be computed using Gaussian elimination:
\begin{equation}
\begin{array}{l@{~}l}
\KKK^1_1 &= [\JJJ^1_1]^{-1} \\
\KKK^2_2 &= [\JJJ^2_2]^{-1} \\
\KKK^1_2 &= - \KKK^2_2 \JJJ^1_2 \KKK^1_1\\
\KKK^3_3 &= [\JJJ^3_3]^{-1} \\
\KKK^2_3 &= - \KKK^3_3 \JJJ^2_3 \KKK^2_2\\
\KKK^1_3 &= - \KKK^3_3 [ \JJJ^1_3 \KKK^1_1 + \JJJ^2_3 \KKK^1_2  ]\\
\cdots
\end{array}
\end{equation}
From this, the GN or LM update steps may be derived directly.

\subsection{Peeling}
\label{sec:peeling}

The \emph{peeling} procedure was originally suggested by \citet{JEN:peeling} as a ``kludge'', i.e. an implementation of DD calibration 
using the DI functionality of existing packages. In a nutshell, this procedure solves for DD gains towards one source at a time, from
brighter to fainter, by

\begin{enumerate}
\item Rephasing the visibilities to place the source at phase centre;
\item Averaging over some time/frequency interval (to suppress the contribution of other sources);
\item Doing a standard solution for DI gains (which approximates the DD gains towards the source);
\item Subtracting the source from the visibilities using the obtained solutions;
\item Repeating the procedure for the next source.
\end{enumerate}

The term ``peeling'' comes from step (iv), since sources are ``peeled'' away one at a time\footnote{The term ``peeling'' has 
occasionally been misappropriated to describe other schemes, e.g. simultaneous independent DD gain solutions. We consider this a 
misuse: both the original formulation by \citet{JEN:peeling}, and the word ``peeling'' itself, strongly implies dealing with 
one direction at a time.}.

Within the framework above, peeling can be considered as the ultimate feed forward approach. Peeling is essentially feed-forward by direction, except rather than taking one step over each direction in turn, each direction is iterated to full convergence before moving on to the next direction. The procedure can then be repeated beginning with the brightest source again, since a second cycle tends to improve the solutions. 

\subsection{Exact matrix inversion}

Better approximations to $(\JHJ)^{-1}$ (or a faster exact inverse) may exist. Consider, for example, Fig.~\ref{fig:JHJ}f: 
the matrix consists of four blocks, with the diagonal blocks being trivially invertible, and the off-diagonal blocks having a 
very specific structure. All the approaches discussed in this paper approximate the off-diagonal blocks by zero, and thus 
yield algorithms which converge to the solution via many cheap approximative steps. If a fast way to invert matrices of the 
off-diagonal type (faster than $O(N^3)$, that is) could be found, this could yield calibration algorithms that converge in fewer 
more accurate iterations.

\begin{figure*}
\begin{center}
\includegraphics[width=\textwidth]{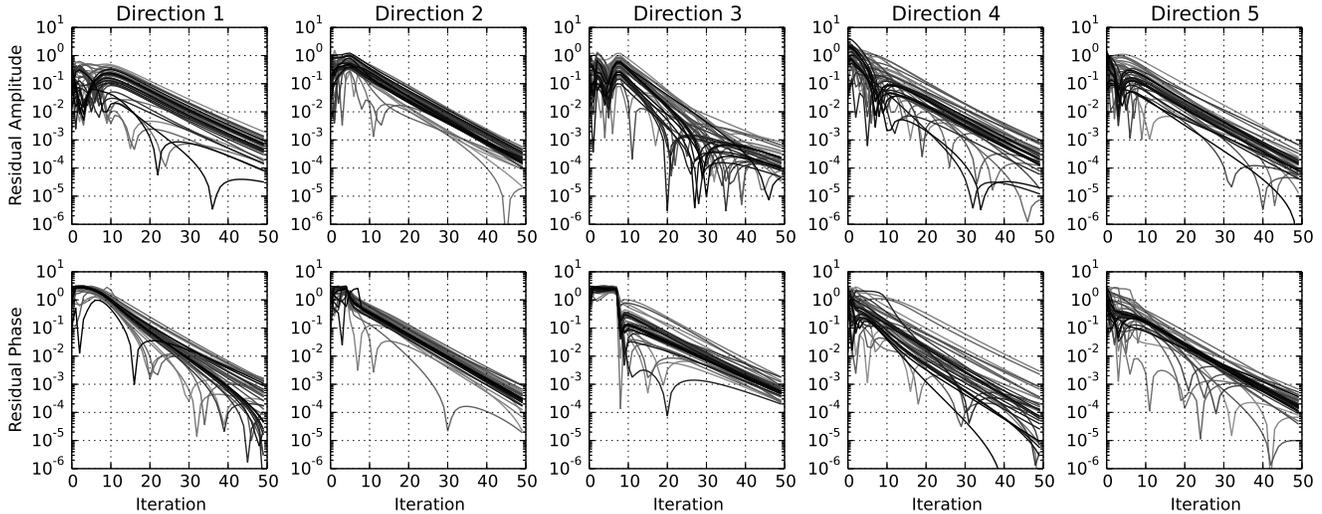}
\caption{\label{fig:Convergence} Amplitude (top row) and phase (bottom row) of the difference between
  the estimated and true gains, as a function of iteration. Columns correspond to directions.
  Different lines correspond to different antennas.}
\end{center}
\end{figure*}

\begin{figure}
\begin{center}
\includegraphics[width=\columnwidth]{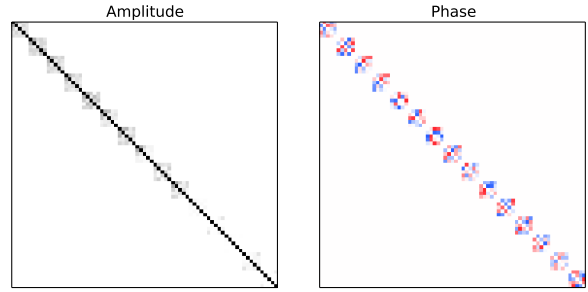}
\caption{\label{fig:HalfJHJ} Amplitude (left panel)
  and phase (right panel) of the block-diagonal matrix $(\JHJ)_\UL$
  for the dataset described in the text. Each block corresponds to one antenna;
  the pixels within a block correspond to directions.}
\end{center}
\end{figure}

\begin{figure*}
\begin{center}
\includegraphics[width=\textwidth]{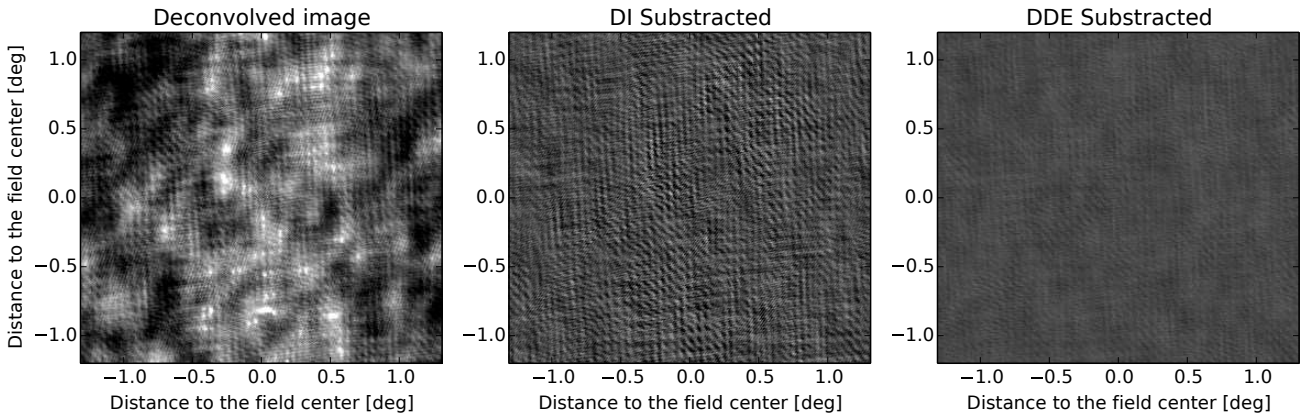}
\caption{\label{fig:resid}Simulation with time-variable DD gains. We show a deconvolved image (left) where
  no DD solutions have been applied, a residual image (centre) made by subtracting the sky model (in the visibility 
  plane) without any DD corrections, and a residual image (right) made by subtracting the sky model with \COH-estimated 
  DD gain solutions (right). The color scale is the same in all panels. In this simulation, applying \COH\ for DD calibration reduces the residual rms level by a factor of $\sim4$.}
\end{center}
\end{figure*}

\begin{figure}
\begin{center}
\includegraphics[width=\columnwidth]{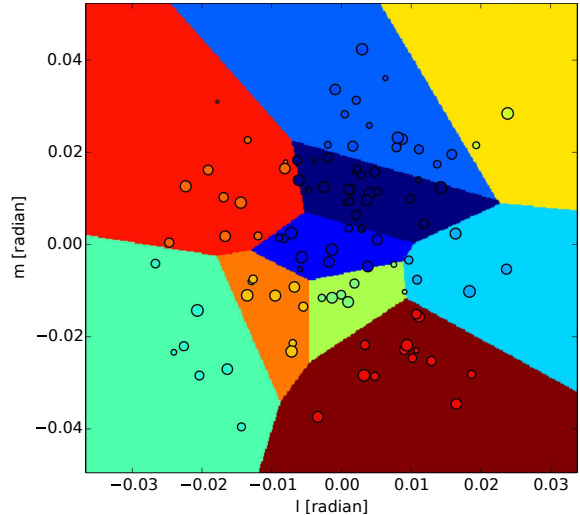}
\caption{\label{fig:tessel} In order to conduct direction-dependent calibration, sources are clustered using a 
Voronoi tessellation algorithm. Each cluster has its own DD gain solution.}
\end{center}
\end{figure}

\section{Implementations}
\label{sec:implementations}

\subsection{\StefCal\ in MeqTrees}

Some of the ideas above have already been implemented in the MeqTrees \citep{meqtrees} version of \StefCal\ 
\citep{OMS-Stefcal}. In particular, the MeqTrees version uses peeling (Sect.~\ref{sec:peeling}) to deal with
DD solutions, and implements fully polarized \StefCal\ with support for both solution intervals and time/frequency smoothing with a Gaussian kernel (as per Sect.~\ref{sec:DI:smooth}). This has already been applied to JVLA L-band data 
to obtain what is (at time of writing) a world record dynamic range (3.2 million) image of the field 
around 3C147 \citep{Perley-3C147}.

\subsection{\COH\ tests with simulated data}

The \COH\ algorithm, in the unpolarized version, has been implemented as a standalone Python script that uses 
the {\tt pyrap}\footnote{{\tt https://code.google.com/p/pyrap}} and 
{\tt casacore}\footnote{{\tt https://code.google.com/p/casacore}} libraries to interface to Measurement Sets. 
This section reports on tests of our implementation with simulated Low Frequency Array (LOFAR) data.

For the tests, we build a dataset using a LOFAR layout with 40 antennas. The phase center is located at
$\delta=+52^\circ$, the observing frequency is set to $50$ MHz
(single channel), and the integrations are 10s. We simulate 20 minutes of data.

For the first test, we use constant direction-dependent gains. We then run \COH\ with a single solution interval 
corresponding to the entire 20 minutes. This scenario is essentially just a test of convergence.
For the second test, we simulate a physically realistic time-variable ionosphere to derive the simulated DD
gains.

\subsubsection{Constant DD gains}
\label{sec:SimpleSimul}

To generate the visibilities for this test, we use a sky model containing five
sources in an ``+'' shape, separated by $1^\circ$. The gains
for each antenna $p$, direction $d$ are
constant in time, and are taken at random along a normal distribution
$g^{(d)}_{p}\sim\mathcal{N}\left(0,1\right)+i\mathcal{N}\left(0,1\right)$. The
data vector $\Dd$ is then built from all baselines, and the full $20$ minutes of data.
The solution interval is set to the full 20 minutes, so a single solution per direction, per antenna is 
obtained.

The corresponding matrix $(\JHJ)_\UL$ is shown in Fig.~\ref{fig:HalfJHJ}. It is block diagonal, each block having size
$\Nd\times\Nd$. The convergence of gain solutions as a function of direction is shown in Fig.~\ref{fig:Convergence}. 
It is important to note that the problem becomes better conditioned (and \COH\ converges faster) 
as the blocks of $(\JHJ)_\UL$ become more diagonally-dominated (equivalently, as the sky model components
become more orthogonal). As discussed in Sect.~\ref{sec:dca}, this happens (i) when more visibilities are taken into account (larger solution intervals) or (ii) if the directions are further away from each other.

\subsubsection{Time-variable DD gains}
\label{sec:VarSimul}

\begin{figure}
\begin{center}
\includegraphics[width=\columnwidth]{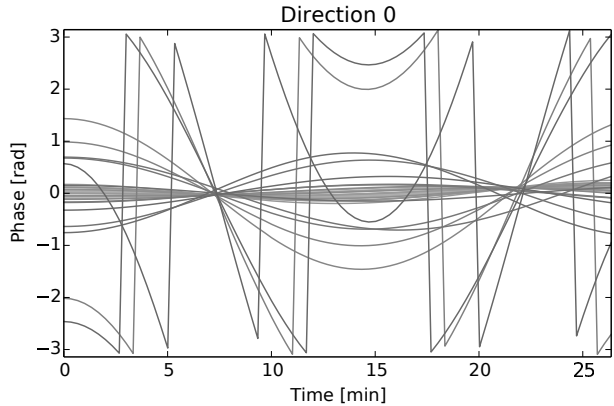}
\caption{\label{fig:PhaseSim} The complex phases of the DD gain terms (for all antennas and a single direction) derived from the 
time-variable TEC screen used in Sect.~\ref{sec:VarSimul}.}
\end{center}
\end{figure}

To simulate a more realistic dataset, we use a sky model composed of 100 point sources of random (uniformly distributed) flux density. 
We also add noise to the visibilities, at a level of about $1\%$ of the total flux. We simulate
(scalar, phase-only) DD gains, using an ionospheric model consisting of a simple phase screen (an infinitesimally thin layer at a 
height of 100 km). The total electron content (TEC) values at the set of sample points are generated using Karhunen-Loeve 
decomposition \citep[the spatial correlation is given by Kolmogorov turbulence, see][]{Tol09}. The constructed TEC-screen has an 
amplitude of $\sim0.07$ TEC-Unit, and the corresponding DD phase terms are plotted in Fig.~\ref{fig:PhaseSim}.

For calibration purposes, the sources are clustered in 10 directions using Voronoi tessellation (Fig.~\ref{fig:tessel}). The solution time-interval is set to $4$ minutes, 
and a separate gain solution is obtained per each direction. Fig. \ref{fig:resid} shows images generated from
the residual visibilities, where the best-fitting model is subtracted in the visibility domain.  The rms residuals
after \COH\ has been applied are a factor of $\sim4$ lower than without DD solutions.

\section*{Conclusions}

Recent developments in optimization theory have extended traditional NLLS optimization approaches
to functions of complex variables. We have applied this to radio interferometric gain calibration, 
and shown that the use of complex Jacobians allow for new insights into the problem, leading
to the formulation of a whole new family of DI and DD calibration algorithms. These algorithms hinge around
different sparse approximations of the $\JHJ$ matrix; we show that some recent algorithmic 
developments, notably \StefCal, naturally fit into this framework as a particular special case
of sparse (specifically, diagonal) approximation. 

The proposed algorithms have different scaling properties depending on the selected matrix approximation
-- in all cases better than the cubic scaling of brute-force GN or LM methods -- and may therefore exhibit 
different computational advantages depending on the dimensionality of the problem (number of antennas,
number of directions). We also demonstrate an implementation of one particular algorithm for DD gain
calibration, \COH.

The use of complex Jacobians results in relatively compact and simple equations, and the resulting algorithms
tend to be embarrassingly parallel, which makes them particularly amenable to implementation on new
massively-parallel computing architectures such as GPUs. 

Complex optimization is applicable to a broader range of problems. Solving for a large number of independent DD gain
parameters is not always desirable, as it potentially makes the problem under-constrained, and can lead to artefacts 
such as ghosts and source suppression. The alternative is solving for DD effect models that employ [a smaller set of]
physical parameters, such as parameters of the primary beam and ionosphere. If these parameters are complex, then
the complex Jacobian approach applies. Finally, although this paper only treats the NLLS problem (thus implicitly
assuming Gaussian statistics), the approach is valid for the general optimization problem as well.

Other approximations or fast ways of inverting the complex $\JHJ$ matrix may exist, and future work can 
potentially yield new and faster algorithms within the same unifying mathematical framework. This flexibility 
is particularly important for addressing the computational needs of the new generation of the 
so-called ``SKA pathfinder'' telescopes, as well as the SKA itself.

\section*{Acknowledgments}

We would like to thank the referee, Johan Hamaker, for extremely valuable comments that improved the paper.
This work is based upon research supported by the South African Research Chairs Initiative of the Department of 
Science and Technology and National Research Foundation. Trienko Grobler originally pointed us in the direction
of Wirtinger derivatives.

\bibliographystyle{mn2e}
\bibliography{cjpaper}

\appendix

\section{$\JJ$ and $\JJ^H\JJ$ for the three-antenna case}
\label{sec:3ant}

To give a specific example of complex Jacobians, consider the 3 antenna case.
Using the numbering convention for $pq$ of 12, 13, 32, we obtain the following partial Jacobians
(Eqs.~\ref{eq:Jk} and \ref{eq:Jbark}):

\begin{equation}
\JJ_k = \Matrix{ccc}{m_{12}\bar{g}_2 & 0 & 0 \\ m_{13}\bar{g}_3 & 0 & 0 \\ 0 & m_{23}\bar{g}_3 & 0 },
\JJ_{k^*} = \Matrix{ccc}{0 & g_1 m_{12} & 0 \\ 0 & 0 & g_1 m_{13} \\ 0 & 0 & g_2 m_{23} }
\end{equation}

We then get the following expression
for the full complex Jacobian $\JJ$ (Eq.~\ref{eq:JJ:di}):

\begin{equation}
\label{eq:JJ:3ant}
\Matrix{cccccc}{
  m_{12}\bar{g}_2 & 0               & 0 &  0          & g_1 m_{12} & 0           \\
  m_{13}\bar{g}_3 & 0               & 0 &  0          & 0          & g_1 m_{13}  \\
  0               & m_{21}\bar{g}_1 & 0 &  g_2 m_{21} & 0          & 0  \\
  0               & m_{23}\bar{g}_3 & 0 &  0          & 0          & g_2 m_{23} \\
  0               & 0               & m_{31}\bar{g}_1 & g_1 m_{31} & 0          & 0  \\
  0               & 0               & m_{32}\bar{g}_2 & 0 & g_3 m_{32} & 0 \\
}
\end{equation}

Then, with the usual shorthand of $y_{pq} = m_{pq} \bar{g}_q$, the
$\JJ^H\JJ$ term becomes:

\begin{equation}
\label{eq:JHJ:3ant}
\Matrix{@{~}c@{}c@{}c@{}c@{}c@{}c@{~}}{
\scriptstyle\yysq{12}{13} &\scriptstyle 0             &\scriptstyle 0             &\scriptstyle 0             &\scriptstyle \bb{1}{2}       &\scriptstyle \bb{1}{3} \\
\scriptstyle0             &\scriptstyle \yysq{12}{23} &\scriptstyle 0             &\scriptstyle \bb{1}{2}       &\scriptstyle 0             &\scriptstyle \bb{2}{3} \\
\scriptstyle0             &\scriptstyle 0             &\scriptstyle \yysq{13}{23} &\scriptstyle \bb{1}{3}       &\scriptstyle \bb{2}{3}       &\scriptstyle 0       \\
\scriptstyle0             &\scriptstyle \bbb{1}{2}      &\scriptstyle \bbb{1}{3}      &\scriptstyle \yysq{12}{13} &\scriptstyle 0             &\scriptstyle 0       \\ 
\scriptstyle\bbb{1}{2}      &\scriptstyle 0             &\scriptstyle \bbb{2}{3}      &\scriptstyle 0             &\scriptstyle \yysq{12}{23} &\scriptstyle 0 \\
\scriptstyle\bbb{1}{3}      &\scriptstyle \bbb{2}{3}      &\scriptstyle 0             &\scriptstyle 0             &\scriptstyle 0             &\scriptstyle  \yysq{13}{23} \\
}
\end{equation}

Finally, the 3-antenna $\JJ^H\Rr$ term becomes

\begin{equation}
\label{eq:JHR:3ant}
\JJ^H\Rr = \Matrix{c}{
\bar{y}_{12} r_{12} + \bar{y}_{13} r_{13} \\
\bar{y}_{21} r_{21} + \bar{y}_{23} r_{23} \\
\bar{y}_{31} r_{31} + \bar{y}_{32} r_{32} \\
y_{12} \bar{r}_{12} + y_{13} \bar{r}_{13}   \\
y_{21} \bar{r}_{21} + y_{23} \bar{r}_{23}   \\
y_{31} \bar{r}_{31} + y_{32} \bar{r}_{32}   \\
}.
\end{equation}

\newcommand{\WW}{\mathbb{W}}
\newcommand{\WWi}{\mathbb{W}^{-1}}

\section{Operator calculus}
\label{sec:opcalculus}

First, let us introduce the vectorization operator ``vec'' and its inverse in the usual (stacked columns) 
way\footnote{Note that \citet{ME4} employs a similar formalism, but uses the (non-canonical) stacked rows definition instead.}. For a $2\times2$ matrix $\mat{X}$:
\begin{equation}
\VEC{\mat{X}} = \Matrix{c}{x_{11}\\x_{21}\\x_{12}\\x_{22}},~~~
\VECINV{\Matrix{c}{x_{11}\\x_{21}\\x_{12}\\x_{22}}} = \mat{X},
\end{equation}
which sets up an isomorphism between the space of $2\times2$ complex matrices $\COMPLEX^{2\times2}$ and the space 
$\COMPLEX^4$. Note that the ``vec'' operator is linear, in other words the isomorphism preserves linear structure:
\begin{equation}
\VEC{(\mat{X}+a\mat{Y})}=\VEC{\mat{X}}+a\,\VEC{\mat{Y}},
\end{equation}
as well as the Frobenius norm:
\begin{equation}
||\VEC{\mat{X}}||_F = ||\mat{X}||_F.
\end{equation}

Consider now the set of all linear operators on $\COMPLEX^{2\times2}$, or $\mathrm{Lin}(\COMPLEX^{2\times2},\COMPLEX^{2\times2}).$
Any such linear operator $\mathcal{B}$, whose action we'll write as $\mathcal{B}\mat{X}$, can be associated with a linear operator 
on 4-vectors $\mat{B}\in\mathrm{Lin}(\COMPLEX^4,\COMPLEX^4)$, by defining $\mat{B}$ as
\begin{equation}
\label{eq:Bx}
\mat{B}\bmath{x} = \VEC{\mathcal{B}\mat{X}},~~~\mat{X}=\VECINV{\bmath{x}}.
\end{equation}
Conversely, any linear operator on 4-vectors $\mat{B}$ can be associated with a linear operator on $2\times2$ matrices by defining
\begin{equation}
\label{eq:BX}
\mathcal{B}\mat{X} = \VECINV{\mat{B}\bmath{x}},~~~\bmath{x}=\VEC{\mat{X}}.
\end{equation}
Now, the set $\mathrm{Lin}(\COMPLEX^4,\COMPLEX^4)$ is simply the set of all $4\times4$ matrix multipliers. Equations~\ref{eq:BX} 
and \ref{eq:BX} establish a one-to-one mapping between this set and the set of linear operators on $2\times2$ 
matrices. In other words, the ``vec'' operator induces two isomorphisms: one between $\COMPLEX^4$ and $\COMPLEX^{2\times2}$,
and the other between $\COMPLEX^{4\times4}$ and linear operators on $\COMPLEX^{2\times2}$. We will designate the second isomorphism 
by the symbol $\WW$:
\begin{equation}
\begin{array}{ll}
\WW\mathcal{B} = \mat{B}: & \mat{B}\bmath{x} = \VEC{\,(\mathcal{B}\,\VECINV{\bmath{x}}}) \\
\WWi\mat{B} = \mathcal{B}: & \mathcal{B}\mat{X} = \VECINV{(\mat{B}\,\VEC{\mat{X}})} 
\end{array}
\end{equation}
Note that $\mathbb{W}$ also preserves linear structure.

Of particular interest to us are two linear operators on $\COMPLEX^{2\times2}$: 
right-multiply by some $2\times2$ complex matrix $\mat{A}$,  and left-multiply by $\mat{A}$:
\begin{equation}
\Rop{\mat{A}}\mat{X} = \mat{XA}, ~~~
\Lop{\mat{A}}\mat{X} = \mat{AX} 
\end{equation}
The $\mathbb{W}$ isomorphism ensures that these operators can be represented as multiplication of 4-vectors 
by specific kinds of $4\times4$ matrices.  The matrix outer product proves to be useful here, and in particular the following 
basic relation:
\begin{equation}
\VEC(\mat{A}\mat{B}\mat{C}) = (\mat{C}^T \otimes \mat{A})\,\VEC{\mat{B}},
\end{equation}
from which we can derive the matrix operator forms via:
\begin{equation}
\begin{array}{l}
\VEC(\II\mat{X}\mat{A}) = (\mat{A}^T \otimes \II)\,\VEC{\mat{X}}\\
\VEC(\mat{A}\mat{X}\II) = (\II \otimes \mat{A})\,\VEC{\mat{X}},
\end{array}
\end{equation}
which gives us
\begin{equation}
\label{eq:RA4x4}
\WW\Rop{\mat{A}} = 
\Matrix{cccc}{a_{11}&0&a_{21}&0 \\ 0&a_{11}&0&a_{21} \\ a_{12}&0&a_{22}&0  \\ 0&a_{12}&0&a_{22} }
\end{equation}
and
\begin{equation}
\label{eq:LA4x4}
\WW\Lop{\mat{A}} = 
\Matrix{cccc}{a_{11}&a_{12}&0&0 \\ a_{21}&a_{22}&0&0 \\ 0&0&a_{11}&a_{12} \\ 0&0&a_{21}&a_{22} }
\end{equation}

From this we get the important property that
\begin{equation}
\label{eq:RH:LH}
[\WW\Rop{\mat{A}}]^H = \WW\Rop{\mat{A}^H},~~~[\WW\Lop{\mat{A}}]^H = \WW\Lop{\mat{A}^H}.
\end{equation}

Note that Eq.~\ref{eq:RAB:LAB}, which we earlier derived from the operator definitions, can now
be verified with the $4\times4$ forms. Note also that Eq.~\ref{eq:RAB:LAB} is equally valid whether 
interpreted in terms of chaining operators, or multipying the equivalent $4\times4$ matrices.


\subsection{Derivative operators and Jacobians}

Consider a matrix-valued function of a a matrix argument and its Hermitian transpose,
$\mat{F}(\mat{G},\mat{G}^H)$. Yet again, we can employ the ``vec'' operator to construct a one-to-one
mapping between such functions and 4-vector valued functions of 4-vectors:
\begin{equation}
\bmath{f}(\bmath{g},\bmath{\bar{g}}) = \VEC{\mat{F}(\VECINV{\bmath{g}},(\VECINV{\bmath{\bar{g}}})^T)}.
\end{equation}

Consider now the partial and conjugate partial Jacobians of $\bmath{f}$ with respect to $\bmath{g}$ 
and $\bar{\bmath{g}}$, defined as per the formalism of Sect.~\ref{sec:Wirtinger}.  
These are $4\times4$ matrices, as given by Eq.~\ref{eq:Jk},
\begin{equation}
\label{eq:Jk-fg}
\JJ_k = [ \partial f_i / \partial g_j ],~~~\JJ_{k^*} = [ \partial f_i / \partial \bar{g}_j ],
\end{equation}
that represent local linear approximations to $\bmath{f}$, i.e. linear operators on $\COMPLEX^4$ that
map increments in the arguments $\Delta\bmath{g}$ and $\Delta\bmath{\bar{g}}$ to increments 
in the function value $\Delta\bmath{f}$. The $\WW$ isomorphism defined above matches these operators
to linear operators on $\COMPLEX^{2\times2}$ that represent linear approximations to 
$\bmath{F}$. It is the latter operators that we shall call \emph{the Wirtinger matrix derivatives}
of $\mat{F}$ with respect to $\mat{G}$ and $\mat{G^H}$:
\begin{equation}
\label{eq:def:dF:dG}
\frac{\partial\mat{F}}{\partial{\mat{G}}} = \WWi(\JJ_k),~~~
\frac{\partial\mat{F}}{\partial{\mat{G}^H}} = \WWi(\JJ_{k^*}^T).
\end{equation}

This is more than just a formal definition: thanks to the $\WW$ isomorphism, the operators 
given by Eq.~\ref{eq:def:dF:dG} are Wirtinger derivatives in exactly the same sense that the Jacobians of 
Eq.~\ref{eq:Jk-fg} are Wirtinger derivatives, with the former being simply the $\COMPLEX^{2\times2}$ manifestation of
the gradient operators in $\COMPLEX^4$, as defined in Sect.~\ref{sec:Wirtinger}. However, operating in 
$\COMPLEX^{2\times2}$ space allows us to write the larger Jacobians of Sect.~\ref{sec:pol} 
in terms of simpler matrices composed of operators, resulting in a Jacobian structure that is entirely analogous
to the scalar derivation.

\section{Gradient-based optimization algorithms}
\label{sec:algs}

This appendix documents the various standard least-squares optimization algorithms 
that are referenced in this paper:

\subsection{Algorithm SD (steepest descent)}

\begin{enumerate}
\item Start with a best guess for the parameter vector, $\bmath{z}_0$;
\item At each step $k$, compute the residuals $\Rr_k$, and the Jacobian
$\JJ=\JJ(\Zz_k)$;
\item Compute the parameter update as (note that due to redundancy, only the top half of the vector actually needs
to be computed):
\begin{equation}
\delta\Zz_k = - \lambda \JJ^H \Rr_k,
\end{equation}
where $\lambda$ is some small value;
\item If not converged\footnote{see below}, set $\bmath{z}_{k+1}=\bmath{z}_k+\delta\zz$, and go back to step (ii).
\end{enumerate}

\subsection{Algorithm GN (Gauss-Newton)}

\begin{enumerate}
\item Start with a best guess for the parameter vector, $\bmath{z}_0$;
\item At each step $k$, compute the residuals $\Rr_k$, and the Jacobian
$\JJ=\JJ(\Zz_k)$;
\item Compute the parameter update $\delta\Zz$ using Eq.~\ref{eq:LM} with $\lambda=0$ (note that only the top half of the vector actually needs to be computed);
\item If not converged, set $\bmath{z}_{k+1}=\bmath{z}_k+\delta\zz$, and go back to step (ii).
\end{enumerate}

\subsection{Algorithm LM (Levenberg-Marquardt)}

Several variations of this exist, but a typical one is:

\begin{enumerate}
\item Start with a best guess for the parameter vector, $\bmath{z}_0$, and an initial value
for the damping parameter, e.g. $\lambda=1$;
\item At each step $k$, compute the residuals $\Rr_k$, and the cost function $\chi^2_k=||\Rr_k||_F$.
\item If $\chi^2_k\ge\chi^2_{k-1}$ (unsuccessful step), reset $\zz_k=\zz_{k-1}$, and set $\lambda=\lambda K$ (where typically $K=10$);
\item Otherwise (successful step) set $\lambda=\lambda/K$;
\item Compute the Jacobian $\JJ=\JJ(\Zz_k)$;
\item Compute the parameter update $\delta\Zz_k$ using Eq.~\ref{eq:LM} (note that only the top half of the vector actually needs to be computed);
\item If not converged, set $\bmath{z}_{k+1}=\bmath{z}_k+\delta\zz$, and go back to step (ii).
\end{enumerate}

\subsection{Convergence}

All of the above algortihms iterate to ``convergence''. One or more of the following convergence criteria may be implemented in each case:

\begin{itemize}
\item Parameter update smaller than some pre-defined threshold: $||\delta \bmath{z}||_F<\delta_0$.
\item Improvement to cost function smaller than some pre-defined threshold: $\chi^2_{k-1}-\chi^2_{k}<\epsilon_0$.
\item Norm of the gradient smaller than some threshold:  $||\JJ||_F<\gamma_0$.
\end{itemize}

\label{lastpage}

\end{document}